\documentclass[reprint,aps,pre,showkeys] {revtex4-1}
\usepackage{bbm}
\usepackage{verbatim}
\usepackage{amsmath}
\usepackage{amssymb}
\usepackage{graphicx}
\usepackage{mathtools}
\usepackage{subfigure}
\usepackage{amsfonts}
\usepackage{hyphenat}
\graphicspath{{./schematics/}{./plots/}}

\newcommand{\qed}{\nobreak \ifvmode \relax \else
      \ifdim\lastskip<1.5em \hskip-\lastskip
      \hskip1.5em plus0em minus0.5em \fi \nobreak
      \vrule height0.75em width0.5em depth0.25em\fi}

\begin{document}

\title{Outbreak statistics and scaling laws for externally driven epidemics}
\author{Sarabjeet Singh}
\email{ss2365@cornell.edu}
\affiliation{Theoretical And Applied Mechanics, Sibley School of Mechanical and Aerospace Engineering, Cornell University, Ithaca, New York}
\author{Christopher R. Myers}
\email{c.myers@cornell.edu}
\affiliation{Laboratory of Atomic and Solid State Physics, and Institute of Biotechnology, Cornell University, Ithaca, New York}
\date{\today}

\begin{abstract}
Power-law scalings are ubiquitous to physical phenomena undergoing a continuous phase transition. The classic Susceptible-Infectious-Recovered (SIR) model of epidemics is one such example where the scaling behavior near a critical point has been studied extensively. In this system the distribution of outbreak sizes scales as $P(n) \sim n^{-3/2}$ at the critical point as the system size $N$ becomes infinite. The finite-size scaling laws for the outbreak size and duration are also well understood and characterized. In this work, we report scaling laws for a model with SIR structure coupled with a constant force of infection per susceptible, akin to a `reservoir forcing'. We find that the statistics of outbreaks in this system are fundamentally different than those in a simple SIR model. Instead of fixed exponents, all scaling laws exhibit tunable exponents parameterized by the dimensionless rate of external forcing.  As the external driving rate approaches a critical value, the scale of the average outbreak size converges to that of the maximal size, and above the critical point, the scaling laws bifurcate into two regimes.  Whereas a simple SIR process can only exhibit outbreaks of size $\mathcal{O}(N^{1/3})$ and $\mathcal{O}(N)$ depending on whether the system is at or above the epidemic threshold, a driven SIR process can exhibit a richer spectrum of outbreak sizes that scale as $O(N^{\xi})$ where $\xi \in (0,1] \backslash \{2/3\}$ and $\mathcal{O}((N/\log N)^{2/3})$ at the multi-critical point.
\end{abstract}


\maketitle

\section{Introduction}
Epidemic models have proven to be extremely useful in understanding the spread of infectious diseases, rumors, computer viruses and fads \cite{andersson2000stochastic,ben2012scaling}. These models constitute a broader category of models describing physical processes that exhibit a second-order phase transition at a critical threshold \cite{sethna2006statistical}. As is characteristic of such transitions, epidemic models exhibit power-law scaling in various statistics characterizing infectious outbreaks at the critical threshold. The classic Susceptible-Infectious-Recovered (SIR) model (eq.\ \ref{eq:SIR}) has been the most widely studied epidemic model in the literature \cite{ben2012scaling,brauer2008mathematical}. In this model, a small number of infected hosts start an `outbreak' in a susceptible pool. An infected host continues to infect susceptible hosts before becoming recovered. The model can be specified using the following rate equations where the rates represent probabilities per unit time of each `reaction' taking place:
\begin{alignat}{2} \label{eq:SIR}
	(S,I,R) &\xrightarrow{\alpha S I / N} &(S-1,I+1,R) \notag\\
	(S,I,R) &\xrightarrow{\quad I \quad} &(S,I-1,R+1) 
\end{alignat}
Note that the rates reported in eq.\ \ref{eq:SIR} are rescaled by the rate of recovery, without loss of generality. 

The SIR model has an epidemic threshold ($\alpha \!=\! 1$ in our case), below which all outbreaks are small (with size $o(N)$) and above which some outbreaks are large (with size $\mathcal{O}(N)$) \cite{brauer2008mathematical}. At the critical threshold, the distribution of outbreak sizes shows the universal scaling of $P(n) \sim n^{-3/2}$ which is invariant to changes in the microscopic details of the model \cite{ben2012scaling,antal2012outbreak}. The size of the average and the maximal outbreaks scale as $N^{1/3}$ and $N^{2/3}$ at the critical point, respectively. \cite{ben2012scaling}. Extensions of the SIR that include multiple stages exhibit the scaling $P(n) \sim n^{-(1 + 2^{-p})}$ for outbreak sizes where $p$ is the number of stages that an infected host crosses before being recovered \cite{antal2012outbreak}. The scaling exponents for the average and the maximal outbreak sizes in this multi-stage SIR are functions of both $p$ and $2^{-p}$, introducing a discrete degree of variability in the scaling depending on the number of stages involved.

\begin{figure}[tbp]
\centering
\includegraphics[width=\columnwidth]{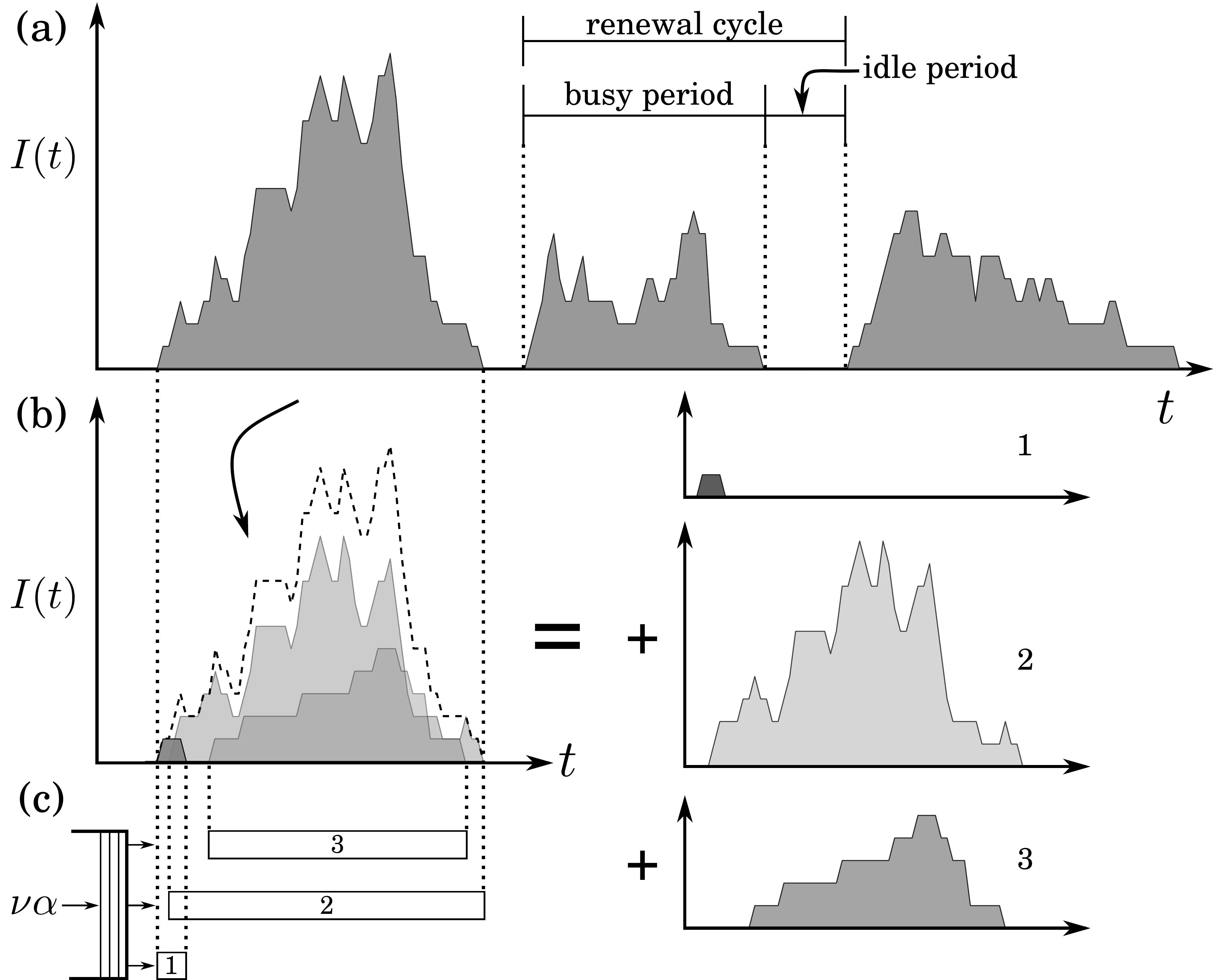}
\vspace{-.15in}
\caption{\label{fig:schematic} (a) A realization of the externally driven SIR process with 3 outbreaks. (b) The decomposition of an outbreak into its constituent micro-outbreaks. Each micro-outbreak is associated with a single imported infection. (c) The $M/G/\infty$ queue where a micro-outbreak is analogous to a customer being serviced at a station. In this realization, the busy period starts when the first customer enters service and ends when the second customer leaves service. The number of customers served during a busy period corresponds to the number of micro-outbreaks that constitute an outbreak.  The statistics of the composite outbreaks depends strongly on the driving rate $\nu$, which is our primary focus here.}
\end{figure}

A different extension to the simple SIR includes an external force of infection action on each susceptible:
\begin{align} \label{eq:SIR_imm}
	(S,I,R) &\xrightarrow{\alpha S (I + \nu) / N} (S-1,I+1,R) \notag\\
	(S,I,R) &\xrightarrow{\quad \; \; \; I \quad \quad} (S,I-1,R+1) 
\end{align}
In this system, each susceptible experiences an additional force of infection with rate $\alpha \nu / N$ where $\nu$ is a dimensionless parameter reflecting the external driving rate. Such a model describes infection dynamics where a pathogen that is sustained in a reservoir repeatedly jumps to susceptible hosts \cite{lloyd2009epidemic,rohani2009environmental}, as might be applicable to the study of cross-species infections such as zoonotic diseases that jump from animals to humans. By construction, the model allows for re-introduction of infection after an outbreak has died out so long as there remain any susceptible host. In the subcritical case ($\alpha < 1$), the process alternates between periods of highly stochastic externally forced outbreaks and periods of no activity (see figure \ref{fig:schematic}a). While the dynamics of this type of model have been examined previously, the calculation of the distributions of outbreak sizes and durations has surprisingly received no attention. These statistics are important for several reasons. From a theoretical perspective, we demonstrate here that outbreak statistics are qualitatively different for the externally driven system than for the simple SIR. From a practical point of view, reservoir-driven zoonotic outbreaks are known to be sporadic \cite{lloyd2009epidemic}, and time series data for outbreaks exhibit active and non-active phases over long periods of time. The information on statistics of individual outbreaks allows one to assess from given data whether the rate $\nu \alpha$ of external introduction and the rate $\alpha$ of infectious contact are constant over a given period of time or varying from outbreak to outbreak. In this work, we solve for the distribution of outbreak sizes $P(n)$ for this system in the limit of $N \to \infty$. From the analytical distribution, we distill scaling laws for all quantities of interest for both infinite and finite population systems. In previous related work, we have calculated similar sorts of outbreak statistics for a different type of externally driven SIR system \cite{singh2013structure}, arising from the coupling of an epidemic outbreak across two populations (e.g., animals and humans).  The case of constant external forcing considered here would be more applicable to situations where infection is at an endemic equilibrium in the reservoir.

\section{Infinite population}
In the limit of infinite system size, the simple SIR process converges in distribution to a linear birth-death (BD) process whose analysis has provided crucial insights in to the full nonlinear process. Similarly, the distribution of the driven SIR process converges to a linear birth-death-immigration (BDI) process (eq.\ \ref{eq:BDI}) as $N \to \infty$, which has been analyzed 
extensively in the literature \cite{zubkov1972life,bailey1990elements}:
\begin{align} \label{eq:BDI}
	(I,R) &\xrightarrow{\alpha(\nu \,+\, I) } (I+1,R) \notag\\
	(I,R) &\xrightarrow{\quad \; I \quad \,} (I-1,R+1) 
\end{align}
Note that the BDI process with $\nu=0$ is identical to the BD process.
Of particular interest here are the sub-critical and critical cases ($\alpha \le 1$)  where outbreaks occur sporadically from imported infections that arrive with rate $\nu \alpha$ but go extinct with probability 1. The time-dependent distribution of the number of infected hosts in the strictly sub-critical case ($\alpha < 1$) which starts with no infection is given by a negative binomial distribution \cite{bailey1990elements},
\begin{equation} \label{eq:I_t_n}
	\mathbb{P} \big[I(t) = n \big] = {n\!+\!\nu\!-\!1 \choose n} (1\!-\!\alpha)^\nu \alpha^n \dfrac{\left[1 \!-\! e^{(\alpha-1)t}\right]^n}{\left[1 \!-\! \alpha e^{(\alpha-1)t}\right]^{n+\nu}}
\end{equation}
which is succinctly expressed using a probability generating function (PGF)
\begin{align} \label{eq:A_xt}
	A(x;t) &= \sum \limits_{n=0}^\infty \mathbb{P} \big [I(t) = n \big ] x^n \notag\\
		   &= \left[ \! \dfrac{1 - \alpha}{1 \!-\! \alpha x + \alpha e^{(\alpha-1)t} (x \!-\! 1)} \! \right]^\nu
\end{align}
Due to an infinite susceptible pool and repeated introductions, the epidemic never goes extinct in the sub-critical process and the number of currently infectious hosts converges to a limiting distribution as $t \rightarrow \infty$ \cite{bailey1990elements}, 
\begin{equation} \label{eq:Ax}
	A(x;\infty) = \sum \limits_{n=0}^\infty \mathbb{P} \big [I(\infty) = n \big ] x^n = \left( \dfrac{1 - \alpha}{1 - \alpha x} \right)^\nu
\end{equation}
The limiting sub-critical BDI process can be interpreted as a renewal process where one idle period $(I = 0)$ and one busy period $(I > 0)$ together form a renewal cycle \cite{zubkov1972life}. We shall define an outbreak in the BDI process to be synonymous with the busy period of the renewal cycle. To obtain the distribution of outbreak sizes, we first draw the analogy between the BDI process and the $M/G/\infty$ queue, as has been reported in literature \cite{zubkov1972life,ong1996class}. The notation $M/G/\infty$ describes a queueing process where customers arrive at an infinite server station according to a Poisson process and enter into service immediately. The service time at a server has a general distribution that is specified. In the notation, $M$ stands for Markovian arrival process, $G$ stands for the general service time distribution and $\infty$ stands for the infinite number of servers \cite{kendall1953stochastic}. The busy period for the queue is defined as the time period when at least one customer is still in service.

Each imported infection can be imagined to be an arrival in the infinite server queue. The service through a single server is then analogous to a `micro-outbreak' in the BDI process, i.e., the chain of infections originating from a single imported infection. A micro-outbreak is then mathematically equivalent to a BD process (eq.\ \ref{eq:BDI} with $\nu=0$) with a single infectious host at the beginning. Thus, the distribution of service times at a single server in the queue system is the same as the distribution of outbreak durations in a BD process, whose closed form solution is available \cite{bailey1990elements}. Finally, the busy period of a BDI process is mathematically equivalent to the busy period of an $M/G/\infty$ queue whose statistics can be calculated using established methods in queueing theory \cite{shanbhag1966infinite}. The intuition for infinite servers comes from the fact that because outbreaks are occurring in an infinite susceptible pool, there is no constraint on how many individual micro-outbreaks can be initiated on overlapping timescales. See figure~\ref{fig:schematic} for illustration of the preceding concepts. Before embarking on new calculations, we first report some results from the literature that we can make use of. For instance, from eq.\ \ref{eq:Ax} the probability that the limiting BDI process is in the idle state ($I=0$) is given by
\begin{equation} \label{eq:q_0}
q_0 = (1 - \alpha)^\nu
\end{equation}
This is equal to the probability that the equivalent $M/G/\infty$ queue is in the idle state. From queueing theory \cite{virtamo2005queueing}, we know
\begin{equation} \label{eq:mean_cust}
	\text{Mean \# customers served in busy period} = 1/q_0
\end{equation}
The number of customers served during the busy period of the queue correspond to the number of imported infections in a single outbreak. 
The total outbreak size can be obtained by integrating over all the micro-outbreaks emanating from imported infections. Heuristically, we
known that the average outbreak size for the BD process is given by $(1 - \alpha)^{-1}$. Using this result, we can guess that the average outbreak size for the BDI process would scale as
\begin{equation} \label{eq:avg_outb_heur}
	\langle n \rangle \sim \dfrac{1}{q_0 (1 - \alpha)} = \left(1 - \alpha \right)^{-(1+\nu)}
\end{equation}
Note that the probability of the process being in the idle state (eq.\ \ref{eq:q_0}) is always greater than 0 as long as $\alpha < 1$ and $\nu$ is finite. Thus, a sub-critical BDI process can never be driven into a perpetual busy period, and accordingly there does not exist any critical driving rate.

The average duration of the outbreak can be derived using the theory of renewal processes (see \cite{zubkov1972life} for a more rigorous derivation). Arrivals in the analogous $M/G/\infty$ queue form a Poisson process with rate $\nu \alpha$. A busy period begins when an arrival takes place at the end of an idle period. Thus, the renewal cycle (idle period + busy period) is a thinned Poisson process which occurs with rate $\nu \alpha q_0$ (the original rate multiplied by the probability that the arrival occurs when the cycle is in the idle state). The average duration of a renewal cycle is $1/(\nu \alpha q_0)$ and the average duration of a busy period is a fraction $1 - q_0$ of the cycle duration. Combining these results, the average duration of an outbreak in the BDI process is given by

\begin{equation} \label{eq:avg_dur}
	\langle t \rangle = \dfrac{1 - q_0}{\nu \alpha q_0} = \dfrac{(1 - \alpha)^{-\nu} - 1}{\nu \alpha}
\end{equation}  
 
The critical BDI process ($\alpha=1$) is an interesting analog to the critical BD process for which some results can be derived using the PGF in equation \ref{eq:A_xt}. For instance, the distribution of the number of infectious hosts as a function of time (with no infection at time 0) is generated by
\begin{subequations}
\begin{equation} \label{eq:K_xt}
	K(x,t) = \left[1 - (x - 1)t \right]^{-\nu}
\end{equation}
which does not have a steady state solution. The average number of infectious hosts grows linearly with time
\begin{equation} \label{eq:mean_I}
	\langle I(t) \rangle = \nu t
\end{equation}
and the probability of the process being in the idle state decays with time. 
\begin{equation}
	p_0(t) = (1+t)^{-\nu}
\end{equation}
\end{subequations}
As expected, the average outbreak size (eq.\ \ref{eq:avg_outb_heur}) and duration (eq.\ \ref{eq:avg_dur}) diverge at the critical threshold. 

\begin{figure}[tbp]
\centering
\includegraphics[width=\columnwidth]{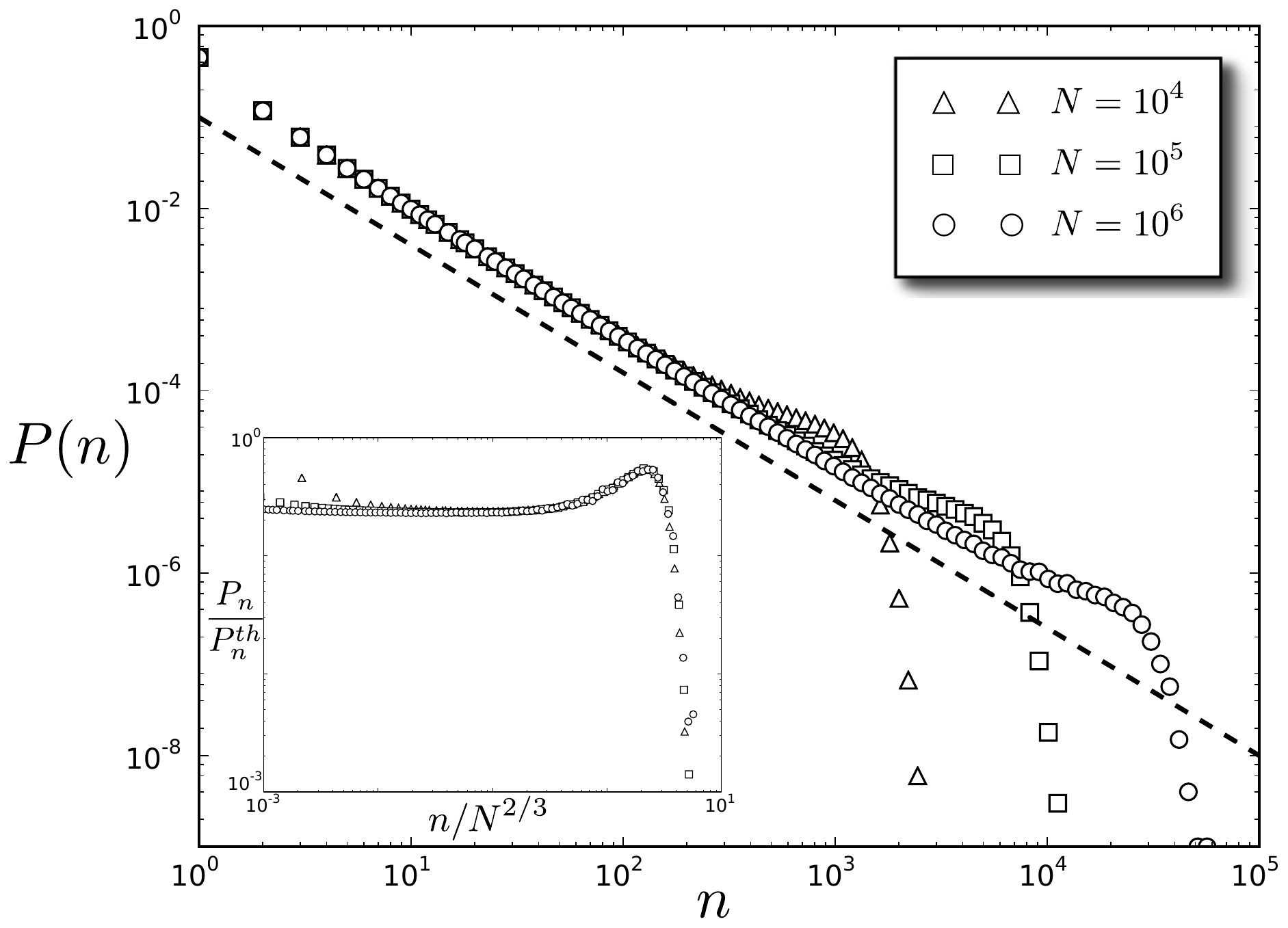}
\vspace{-.15in}
\caption{\label{fig:outb_dist_scaling_nu_lt_1} The probability of having an outbreak of size $n$ empirically calculated from $10^9$ realizations of the process for different values of $N$ and $\alpha = 1, \nu=0.2$. The dashed line shows the slope of the analytical scaling predicted from theory (eq.\ \ref{eq:Pn_nu_lt_1}), ignoring any constant prefactors. Inset shows the collapse of outbreak sizes when scaled by $N^{2/3}$. The bump near the exponential cutoff represents the probability mass associated with outbreaks that would have continued along the power-law in an infinite size system, but are clustered due to finite size effects.}
\end{figure}

The existing generating functions for the BDI process (eq.\ \ref{eq:A_xt} and \ref{eq:K_xt}) do not describe the busy period of the process in isolation. One can only query the distribution of the number of infected (or recovered) hosts at time $t$ without conditioning on whether the process is busy or idle and without any knowledge of how many outbreaks have occurred before $t$. For calculating the statistics of a single outbreak, integrating the time-dependent generating functions unconditionally would be incorrect. Instead, one must integrate over the duration of a single outbreak, which corresponds to the busy period of the analogous $M/G/\infty$ queue. The calculation for the number of customers served in the busy period exists for the $M/G/\infty$ queue \cite{shanbhag1966infinite} that we shall adopt for our purpose. The calculation presented here is done for arbitrary values of $\alpha$ and $\nu$ assuming that the outbreak sizes are finite. The first ingredient in this calculation are the statistics of a BD process, which are summarized in the following PGF. Let
\begin{equation}
	F(x,y;t) \!=\! \sum_{m,n} \mathbb{P} \big [ I(t) \!=\! m, R(t) \!=\! n \big ] \, x^m y^n
\end{equation}
be the PGF for the joint distribution of infectious and removed hosts in a BD process with birth rate $\alpha$ and death rate set to 1 that starts with one infectious individual at time 0. From \cite{bailey1990elements} the exact solution of the PGF is given by
\begin{align} \label{eq:F}
	F(x,y;t) = \dfrac{\Lambda_0(\Lambda_1  - x) + \Lambda_1(x - \Lambda_0)e^{-\alpha (\Lambda_1-\Lambda_0) t}} {(\Lambda_1-x) + (x - \Lambda_0) e^{-\alpha (\Lambda_1-\Lambda_0) t}}
\end{align}
where $\Lambda_0(y)$ and $\Lambda_1(y)$ are roots of the following quadratic equation such that $0 < \Lambda_0 < 1 < \Lambda_1$.
\begin{equation} \label{eq:quad}
\alpha w^2 - \left(\alpha + 1 \right) w + y = 0
\end{equation}
The joint distribution of the duration $T$ and the size $R(T)$ of an outbreak can be summarized using $F(0,y;t)$, i.e.,
\begin{equation} \label{eq:F_0y}
	F(0,y;t) = \sum_{n \ge 1} \mathbb{P} \big[ T \!\le\! t, R(T) \!=\! n \big ] \, y^n
\end{equation}
The trick that yields the desired result is to use the PGF $F(0,y;t)$ in place of the service time distribution for calculating the number of customers served in a busy period of $M/G/\infty$ queue (see Appendix for details). The intuition comes from the fact that the outbreak duration and size are correlated random variables, and integrating the joint distribution preserves the correlation. Once we substitute $F(0,y;t)$ and simplify the integration, we obtain the following PGF (eq.\ \ref{eq:G}) for the joint distribution of the number of imported infections and outbreak size during the busy period of the BDI process.
\begin{subequations}
\begin{equation} \label{eq:G}
	G(x,y) = 1 - \dfrac{1}{\nu}\dfrac{\Lambda_1 \, z ^a \left(1 - z \right)^b}{\int \limits_z^1 r^{a-1} (1 - r)^{b-1} dr}
\end{equation}
where
\begin{align}
z = 1 - \dfrac{\Lambda_0}{\Lambda_1}, &\quad a = 1 - \nu x, \quad b = \nu \left( \dfrac{1 - \Lambda_0 x}{\Lambda_1 - \Lambda_0} \right)  \label{eq:zab_1} \\
\Lambda_0,\Lambda_1 &= \dfrac{(\alpha+1) \mp \sqrt{(\alpha+1)^2 - 4 \alpha y}}{2 \alpha} \label{eq:lambda_1}
\end{align}
\end{subequations}
The marginal distribution of outbreak sizes is generated by $G(1,y)$. Let $H(y)$ be the PGF for the marginal distribution at the critical threshold $\alpha=1$. This simplifies some of the terms in the PGF:
\begin{subequations}
\begin{equation} \label{eq:H}
	H(y) = 1 - \dfrac{1}{\nu}\dfrac{\Lambda_1 \, z ^a \left(1 - z \right)^b}{\int \limits_z^1 r^{a-1} (1 - r)^{b-1} dr}
\end{equation}
where
\begin{align}
z = 1 -& \dfrac{\Lambda_0}{\Lambda_1}, \quad a = 1 - \nu, \quad b = \dfrac{\nu}{2} \label{eq:zab_2} \\
&\Lambda_0,\Lambda_1 = 1 \mp \sqrt{1 - y} \label{eq:lambda_2}
\end{align}
\end{subequations}
The integral in the denominator of eq.\ \ref{eq:H} can be solved explicitly for $\nu \in \mathbb{Z}_{> 0}$. For an arbitrary $\nu$, the integral can be represented as the difference between the Beta function $\mathrm{B}(a,b)$ and incomplete Beta function $\mathrm{B}(a,b;z)$. Let this integral be denoted by $\mathrm{J}(a,b;z)$.
\begin{align}
\mathrm{J}(a,b;z) = \mathrm{B}(a,b) - \mathrm{B}(a,b;z)
\end{align}
The asymptotic form for $P(n)$ -- the probability of having an outbreak of size $n$ -- can be obtained by the singularity expansion of the PGF $H(y)$ around $y=1$ (or $z = 0$). For $\nu < 1$ (which implies $a,b > 0$), the incomplete Beta function can be approximated by 
\begin{equation}
	\mathrm{B}(a,b;z) \sim \dfrac{z^a(1-z)^b}{a}
\end{equation}
in the limit of $z \to 0$ \cite{press1992numerical}. The PGF $H(y)$ simplifies as,
\begin{align}
	&H(y) \sim 1 - \dfrac{1}{\nu \mathrm{B}(a,b)}\dfrac{\Lambda_1 \, z ^a \left(1 - z \right)^b}{\left(1 - \dfrac{z^a(1-z)^b}{a \mathrm{B}(a,b)}\right)} \notag\\
	&= 1 - \dfrac{1}{\nu \mathrm{B}(a,b)} \Lambda_1 \, z ^a \left(1 - z \right)^b \left[1 + \dfrac{z^a(1-z)^b}{a \mathrm{B}(a,b)} + \cdots \right] \notag\\[2pt]
	&= 1 - \dfrac{2^{1-\nu} (1 -y)^{(1-\nu)/2} (y)^{\nu/2}}{\nu \mathrm{B}(a,b)} + \cdots
\end{align}
where the simplification in the last step follows from substituting for $z$ from eq.\ \ref{eq:zab_2}. From the leading order term $(1-y)^{(1-\nu)/2}$, we can assert the following asymptotic form for $P(n)$ as described in \cite{flajolet2009analytic}
\begin{equation} \label{eq:Pn_nu_lt_1}
	P(n) \sim n^{-(3-\nu)/2}
\end{equation}
The scaling law is verified in figure~\ref{fig:outb_dist_scaling_nu_lt_1}. As expected, the power-law becomes more and more flat with increasing $\nu$. The pronounced bump in the simulations is a finite size effect due to the clustering of outbreaks that would have continued to exhibit the power-law scaling if the system size was infinite \cite{christensen2005complexity}. It can be verified that all moments of the distribution diverge for any value of $\nu \in [0,1)$. 

\begin{figure}[tbp]
\centering
\includegraphics[width=\columnwidth]{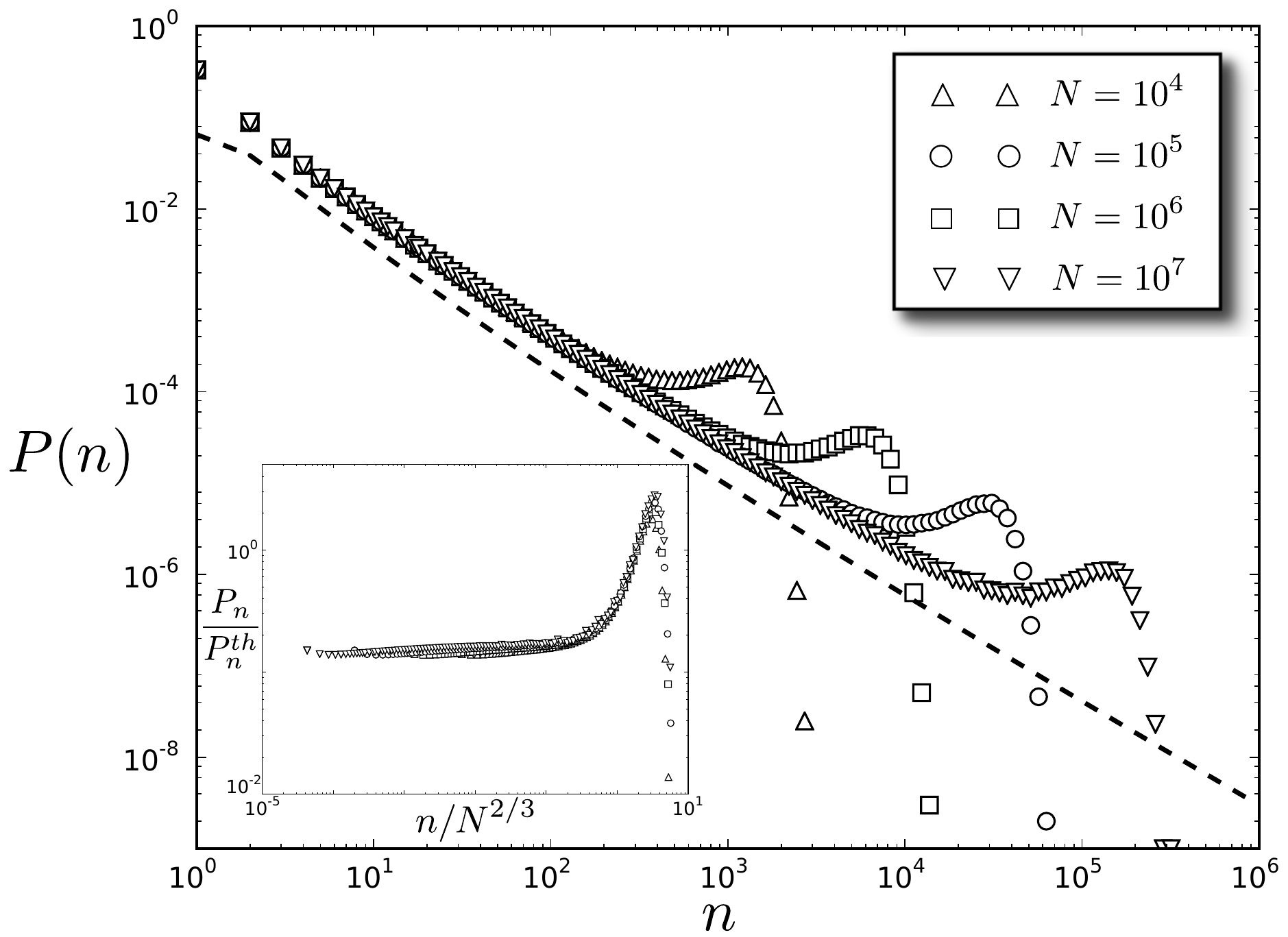}
\vspace{-.15in}
\caption{\label{fig:outb_dist_scaling_nu_eq_1} The probability of having an outbreak of size $n$ for $\alpha = 1, \nu=1$. Dashed line plotted at an offset represents the analytical scaling (eq.\ \ref{eq:Pn_nu_1}). Inset shows the collapse of outbreak sizes when scaled by $N^{2/3}$. }
\end{figure}

At $\nu=1$, the PGF $H(y)$ in eq.\ \ref{eq:H} can be simplified further into a closed form solution,
\begin{align} \label{eq:Hy_nu1}
	H(y) = 1 + \dfrac{\sqrt{y}}{\log \left(\dfrac{\sqrt{1-y}}{1 + \sqrt{y}}\right)}
\end{align}
whose singularity analysis around $y=1$ yields the following asymptotic form for $P(n)$
\begin{equation} \label{eq:Pn_nu_1}
	P(n) \sim \dfrac{1}{n \log^2 4n} \left[\dfrac{1}{2} - \dfrac{\gamma}{\log 4n} + \mathcal{O} \left( \dfrac{1}{\log^2 4n} \right) \right]
\end{equation}
where $\gamma = 0.5772 \ldots$ is the Euler-Mascheroni constant. As in the case of $\nu < 1$, all moments of the distribution diverge in this case as well. See figure~\ref{fig:outb_dist_scaling_nu_eq_1} for comparison with stochastic simulations.

The case of $\nu > 1$ requires a careful analysis of the function $\mathrm{J}(a,b;z)$ because the parameter $a$ becomes negative in this regime. Since $\mathrm{J}(a,b;z)$ is the difference of beta and incomplete beta functions, the following identity holds
\begin{equation} \label{eq:beta_prop}
	\mathrm{J}(a,b;z) = \dfrac{(a+b)}{a} \mathrm{J}(a+1,b;z) - \dfrac{z^a (1-z)^b}{a}
\end{equation}
Consider the case where $\nu \in (1,2)$ which implies that $0 < a+1 < 1$. In this case as $z \to 0$,
\begin{align}
	&\mathrm{J}(a+1,b;z) \sim \mathrm{B}(a+1,b) \notag\\
	&\mathrm{J}(a,b;z) \sim \dfrac{(a+b)}{a} \mathrm{B}(a+1,b;z) - \dfrac{z^a (1-z)^b}{a}
\end{align}
and the PGF $H(y)$ simplifies as follows
\begin{align}
	&H(y) \sim 1 + \dfrac{a}{\nu} \dfrac{\Lambda_1}{\left[ 1 - (a+b) \mathrm{B}(a+1,b) z^{-a} (1-z)^{-b}\right]} \\
		 &= 1 + \dfrac{a \Lambda_1}{\nu} \left[ 1 + (a+b) \mathrm{B}(a+1,b) z^{-a} (1-z)^{-b} + \cdots \right] \notag
\end{align}
Substituting for $z$, we obtain a series expansion in fractional powers of $\sqrt{1-y}$. The leading term in the expansion is of the order $(1-y)^{(\nu-1)/2}$ which provides the following asymptotic form for $P(n)$. 
\begin{align} \label{eq:Pn_nu_gt_1}
	P(n) \sim n^{-(\nu + 1)/2}
\end{align}
\begin{figure}[tbp]
\centering
\includegraphics[width=\columnwidth]{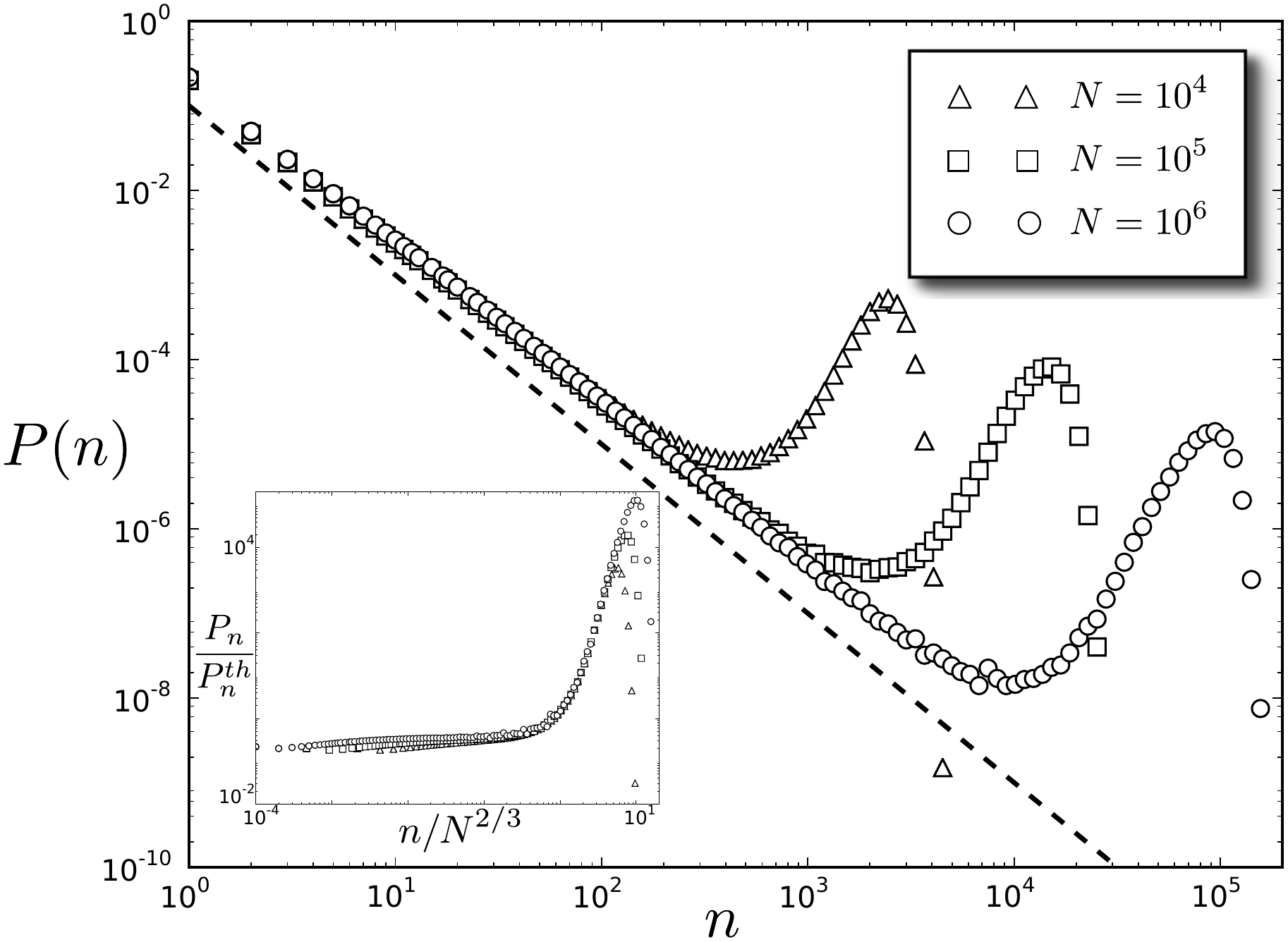}
\vspace{-.15in}
\caption{\label{fig:outb_dist_scaling_nu_gt_1} The probability of having an outbreak of size $n$ for $\alpha = 1, \nu=3$. Dashed line represents the analytical scaling (eq.\ \ref{eq:Pn_nu_gt_1}) at an offset. The finite size bump is more pronounced because it has accumulated outbreaks of two distinct scales. The bump starts at the scale of $N^{2/3}$ consistent with what is observed for $\nu \le 1$. However, the exponential cutoff which marks the end of the bump occurs at a different scale as evident from the lack of scaling collapse for large $n$ in the inset. The presence of two scales in the bump is discussed in detail in section \ref{subsec:outbreak_size} and illustrated in figure~\ref{fig:scaling_collapse_nu_gt_1}.}
\end{figure}
Similarly, by binning $\nu$ in $\{(2,3),(3,4),\cdots\}$ and applying the property given by eq.\ \ref{eq:beta_prop} iteratively, we obtain the asymptotics as in eq.\ \ref{eq:Pn_nu_gt_1}. The same is true when integral values are chosen for $\nu$, in which case the PGF can be simplified further. For instance, substituting $\nu=2$ in eq.\ \ref{eq:H} simplifies the PGF to 
\begin{equation}
	H(y) = \dfrac{1 - \sqrt{1-y}}{2}
\end{equation}
which is same as the PGF for the BD process with a prefactor of $1/2$. Asymptotic analysis reveals scaling exponent of $3/2$ consistent with eq.\ \ref{eq:Pn_nu_gt_1}. The agreement of eq.\ \ref{eq:Pn_nu_gt_1} with stochastic simulations is demonstrated in figure~\ref{fig:outb_dist_scaling_nu_gt_1}. 

Equation \ref{eq:Pn_nu_gt_1} suggests that the outbreak size distribution falls off more steeply with increasing 
$\nu$.  This seems counterintuitive at first because one would expect that, with increasing $\nu$, there should be a greater probability of larger outbreaks, leading to more slowly decaying distribution. The resolution of this puzzle can be found by looking at the total probability mass contained in the generating function. It can be verified that for $\nu \le 1$, $H(y) \rightarrow 1$ as $y \rightarrow 1$, i.e., the distribution is proper. But for $\nu > 1$, the distribution becomes defective such that
\begin{equation}
	\lim_{y \to 1} H(y) = \nu^{-1}
\end{equation}
The remaining probability mass $1 - \nu^{-1}$ is associated with the infinite sized outbreak. This effect can be seen in stochastic simulations (figure~\ref{fig:outb_dist_scaling_nu_gt_1}) where the  outbreaks not accounted by the power-law cluster in the bump of the distribution. More formally, we have the following
\begin{equation}
	\lim_{n \to \infty}	\mathbb{P} \big [R (\infty) > n \big ] \sim \begin{cases}
															\mathcal{O}\left( n^{-(1-\nu)/2} \right)  & \text{$\nu < 1$,} \\[5pt]
\mathcal{O}\left( (\log n)^{-1} \right) & \text{$\nu = 1$,} \\[5pt]															
1 \!-\! \nu^{-1} \!+\! \mathcal{O}\left( n^{-(\nu-1)/2} \right) & \text{$\nu > 1$.}															
														 														\end{cases} 
\end{equation}
Thus, for $\nu > 1$ the probability of having an outbreak size exceeding any arbitrary scale converges to a constant value of $1-\nu^{-1}$ whereas the same probability diminishes with $n$ for the case of $\nu \le 1$. For $\nu > 1$, the distribution represented by the generating function $H(y)$ excludes the infinite sized outbreak. Thus, the statistics are conditional on a finite sized outbreak. The distribution falls more steeply with increasing $\nu$ because more and more outbreaks escape to infinity with probability $1 - \nu^{-1}$. Nevertheless, the statistics of the power-law regime are interesting to analyze even if they represent part of the distribution. For instance, the $k^{th}$ moment of the distribution is finite only if $\nu > 2k+1$ and diverges otherwise.
\begin{equation} \label{eq:power_law_div}
	\langle n^k \rangle \sim \begin{cases}
							\dfrac{1}{\nu - (2k+1)} & \text{if $\nu > 2k+1$}, \\
							\infty & \text{otherwise.}
							\end{cases}
\end{equation}
Finally, we summarize the asymptotic statistics calculated in the preceding section
\begin{equation}
	P(n) \sim \begin{cases}
				n^{-(3-\nu)/2} & \text{$\nu < 1$,} \\[5pt]
			   	\! \dfrac{1}{n \log^2 4n} \left[\dfrac{1}{2} - \dfrac{\gamma}{\log 4n} + \mathcal{O} \left( \dfrac{1}{\log^2 4n} \right) \right]  & \text{$\nu = 1$,} \\[15pt]
			   	n^{-(\nu + 1)/2} & \text{$\nu > 1$.}
			\end{cases}
\end{equation}
We can now put the results in some perspective. The external driving can be thought of as a `coupling agent' that combines an increasing number of micro-outbreaks into a single outbreak as $\nu$ is increased. When $\alpha=1$ and $\nu$ is above 1, the external driving binds an infinite number of micro-outbreaks into one contiguous outbreak with probability $1 - \nu^{-1}$. Qualitatively, the BDI process can also be interpreted as a two-state Markov chain that switches between the  idle period and the busy period. In this interpretation the idle period is positive recurrent if $\alpha < 1$ (busy period ends with probability 1 and in finite time),  null recurrent if $\alpha = 1, \nu \le 1$ (busy period ends with probability 1 but the expected duration is $\infty$) and transient if $\alpha = 1, \nu > 1$ (busy period can persist indefinitely). The case of $\alpha > 1$ is trivial since a supercritical process can grow exponentially even without the external forcing. The idle period is thus transient in this case.

\section{Finite population}
\subsection{Outbreak size} \label{subsec:outbreak_size}
\begin{figure}[tbp]
\centering
\includegraphics[width=\columnwidth]{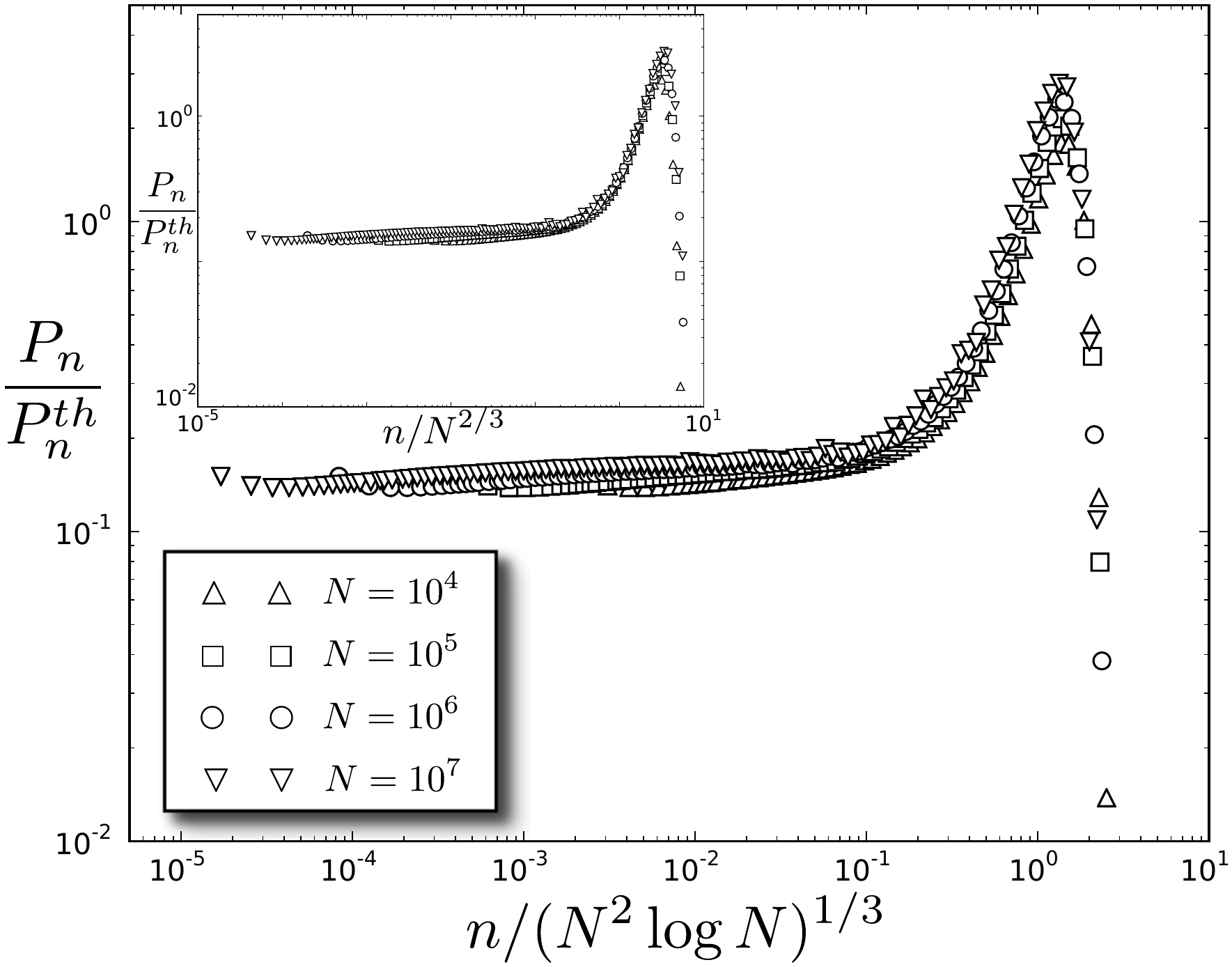}
\vspace{-.15in}
\caption{\label{fig:scaling_collapse_nu_eq_1} Scaling collapse at $M \sim (N^2 \log N)^{1/3}$ for $\nu=1$. Y-axis is scaled by the theoretical scaling law of eq.\ \ref{eq:Pn_nu_1}. Note that the scaling collapse is distinct from the one shown in inset (done at the scale of $N^{2/3}$). While the power-laws collapse on top of each other at $N^{2/3}$, the exponential cutoffs collapse at $(N^2 \log N)^{1/3}$. This separation of scales is more pronounced for $\nu > 1$ (see figure~\ref{fig:scaling_collapse_nu_gt_1}).}
\end{figure}

\begin{figure}[tbp]
\centering
\includegraphics[width=\columnwidth]{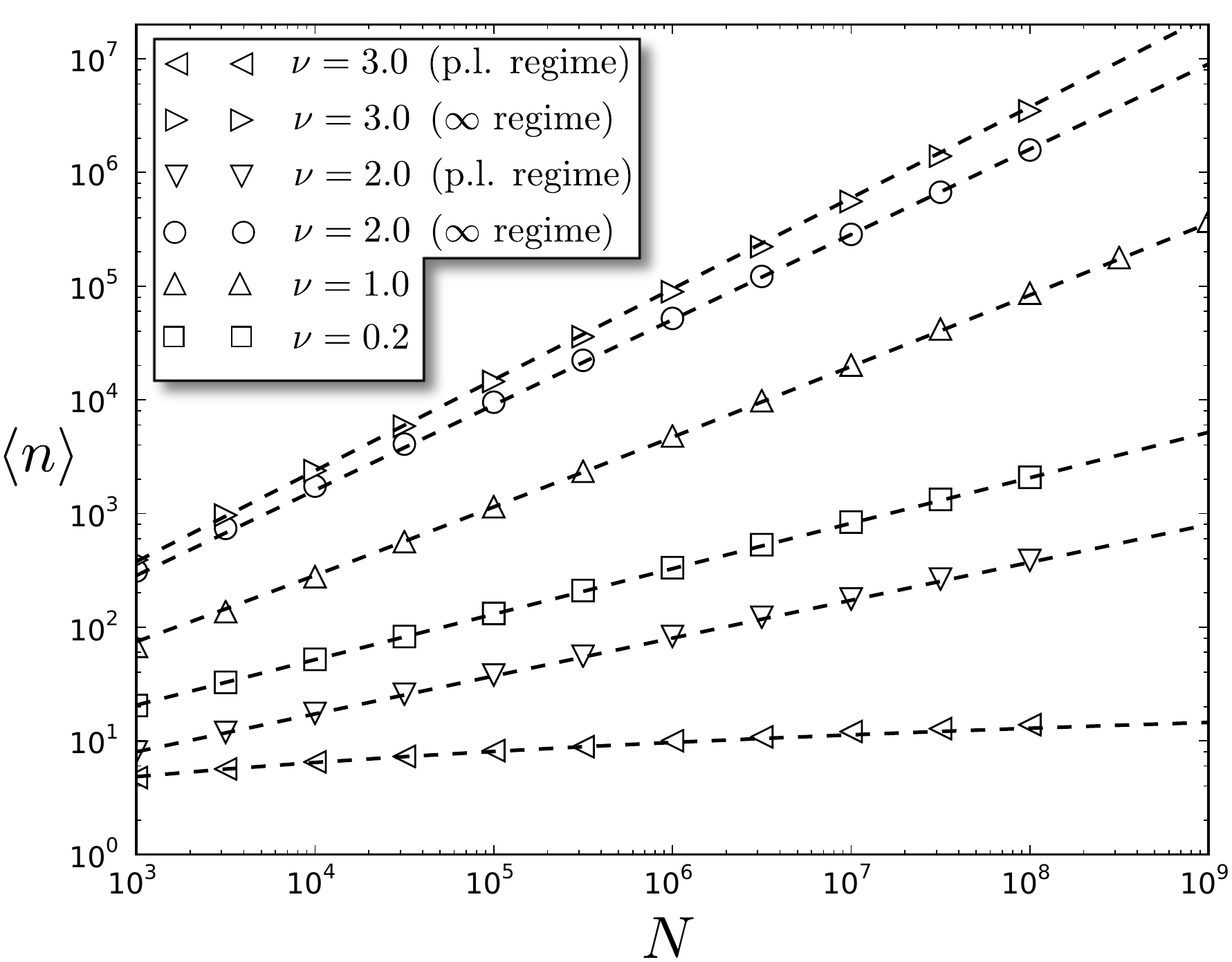}
\vspace{-.15in}
\caption{\label{fig:avg_outb_size} Finite size scaling for average outbreak size ($\nu=\{0.2,1,2,3\}$). The statistics of the power law regime and the infinite regime were calculated separately using $N^{2/3}$ as the separation boundary. Note that the scaling for $
\nu = 3$ is purely logarithmic in the power-law regime. Dashed lines represent the scaling laws predicted from theory (eq.\ \ref{eq:avg_outb_summ}).}
\end{figure}

For a finite-sized system (eq.\ \ref{eq:SIR_imm}), we first establish the scaling of the `maximal' outbreak size echoing the analysis in \cite{ben2012scaling,antal2012outbreak}. Let there be a maximal size $M$, such that the outbreak can not exceed this size due to depletion in the susceptible pool. For $\nu < 1$, the algebraic distribution in eq.\ \ref{eq:Pn_nu_lt_1} gives an estimate for the average outbreak size:
\begin{equation} \label{eq:n_M_rel}
\langle n \rangle = \sum_{n \le M} n \cdot P(n)  \sim \sum_{n \le M} n^{-(1-\nu)/2} \sim M^{(1+\nu)/2}
\end{equation}
In a finite-sized system, the effective rate of infectious contact per infected host is reduced to $\alpha_{\star} = 1 - M/N$ due to depletion. From eq.\ \ref{eq:avg_outb_heur}, we obtain a second estimate for the scaling of the average outbreak size:
\begin{equation} \label{eq:avg_with_M}
	\langle n \rangle \sim \left(1 - \alpha_{\star} \right)^{-(1+\nu)}
	= (N/M)^{1 + \nu}.
\end{equation}
Equating the two estimates we obtain the following scaling laws for $\nu < 1$:
\begin{equation} \label{eq:M1}
	M \sim N^{2/3}, \quad \text{and} \quad \langle n \rangle \sim N^{(1+\nu)/3}
\end{equation}
The scaling of $M$ is verified in figure~\ref{fig:outb_dist_scaling_nu_lt_1} (inset) and that of $\langle n \rangle$ in figure \ref{fig:avg_outb_size}.
For $\nu = 1$, we use the expression for $P(n)$ in eq.\ \ref{eq:Pn_nu_1} and obtain $M$ as the solution of the following implicit equation:
\begin{equation} \label{eq:M_nu_1}
\left( \dfrac{N}{M} \right)^2 =	\dfrac{M}{2 \log^2 4M} + \mathcal{O} \left( \dfrac{M}{\log^3 4M} \right)
\end{equation}
whose solution to a first order approximation leads to the following scaling laws
\begin{equation} \label{eq:M2_th}
	M \sim \left( N \log N \right)^{2/3} \quad \text{and} \quad \langle n \rangle \sim \left(\dfrac{N}{\log^2 N}\right)^{2/3}
\end{equation}
However, numerical results obtained from stochastic simulations reveal slightly different scaling laws
\begin{equation} \label{eq:M2}
	M \sim (N^2 \log N)^{1/3}, \quad \text{and} \quad \langle n \rangle \sim \left(\dfrac{N}{\log N}\right)^{2/3} 
\end{equation}
that differ from theory by a factor of $(\log N)^{1/3}$ in $M$ and $(\log N)^{-2/3}$ in $\langle n \rangle$. The empirical scaling law can be obtained if eq.\ \ref{eq:M_nu_1} is replaced with the following
\begin{equation} \label{eq:M2_scl}
\left(\dfrac{N}{M}\right)^2 \sim \dfrac{M}{\log M}
\end{equation}
Although the power-law part of the scaling -- that is the term $N^{2/3}$ -- is consistent between both the empirically observed (eq.\ \ref{eq:M2}) and the theoretically calculated (eq.\ \ref{eq:M2_th}) scaling, we are unable to resolve the logarithmic corrections and pose their solution as an open problem. The agreement of the scaling laws (eq.\ \ref{eq:M2}) with results from stochastic simulations is shown in figures~\ref{fig:scaling_collapse_nu_eq_1} and \ref{fig:avg_outb_size}. Henceforth, we shall refer only to the empirical scaling law for $\nu=1$ where the logarithmic corrections are important.

The case of $\nu > 1$ requires careful consideration. The analysis on the infinite-sized system revealed that outbreaks occur according to a power law distribution (eq.\ \ref{eq:Pn_nu_gt_1}) with probability $\nu^{-1}$ or are infinite in size with probability $1 - \nu^{-1}$. Henceforth, we shall label these as the `power-law regime' and the `infinite regime', respectively. The average outbreak size in the power-law regime diverges when $\nu < 3$ (see eq.\ \ref{eq:power_law_div}). For finite systems, we expect that both the infinite regime and the power-law regime would admit two different scaling laws for average outbreak size and duration. The power-law regime admits a positive exponent for the scaling law only for $\nu \in (1,3)$. Let $\langle n \rangle_\infty$ be the average outbreak size conditioned on the outbreak being in the infinite regime. From eq.\ \ref{eq:n_M_rel}, note that as $\nu \to 1$, the average outbreak size $\langle n \rangle$ approaches $M$ in scale. For $\nu = 1$, we found empirically that $\langle n \rangle \sim M /\log M$ (see eq.\ \ref{eq:M2_scl}). For $\nu > 1$, intuition suggests that in the infinite regime $\langle n  \rangle_{\infty} \sim M$, i.e., all outbreaks will be clustered at one scale. Using eq.\ \ref{eq:avg_with_M}, we obtain the following scaling relationship
\begin{equation} \label{eq:M3}
	\langle n \rangle_{\infty} \sim M \sim N^{(\nu+1)/(\nu+2)}
\end{equation}

\begin{figure}[tbp]
\centering
\includegraphics[width=\columnwidth]{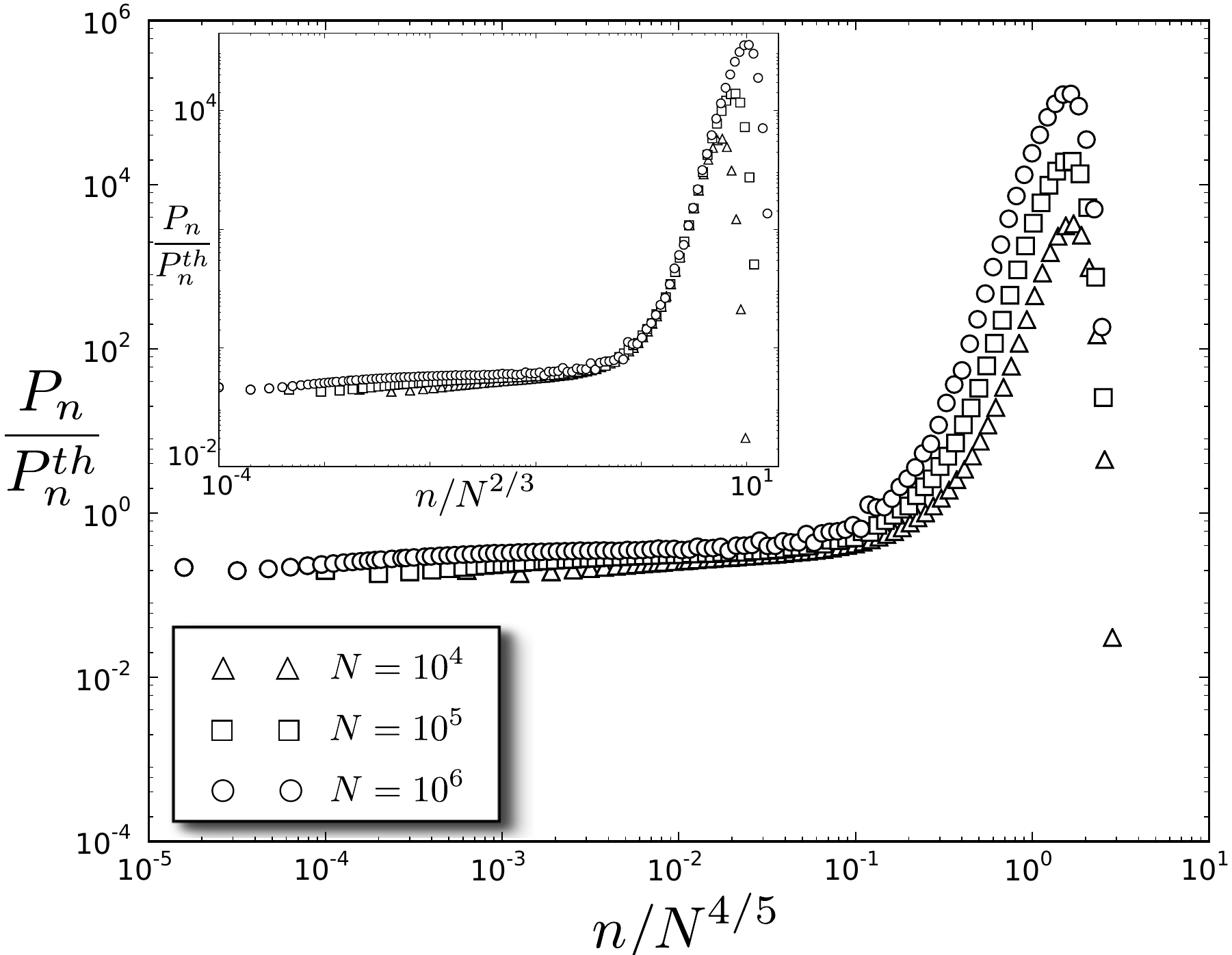}
\vspace{-.15in}
\caption{\label{fig:scaling_collapse_nu_gt_1} Scaling of the outbreak size distribution by $M \sim N^{(\nu+1)/(\nu+2)}$ for $\nu = 3$. The y-axis is scaled by the theoretical scaling law of eq.\ \ref{eq:Pn_nu_gt_1}. Similar to figure~\ref{fig:scaling_collapse_nu_eq_1}, the scaling collapse is distinct from the one shown in inset (outbreaks scaled by $N^{2/3}$). The power-law regime exhibit a scaling collapse at $N^{2/3}$, while the exponential cutoffs collapse at $N^{4/5}$.}
\end{figure}

The exponent of the scaling law in eq.\ \ref{eq:M3} is an increasing function of $\nu$ that lies in the interval $(2/3,1)$ for all $\nu > 1$. The lower bound of $2/3$ is consistent with the fact that the scaling law for $\nu < 1$ has $2/3$ as the upper bound (eq.\ \ref{eq:M1}) and that the same exponent shows up at $\nu=1$ albeit with logarithmic factors (eq.\ \ref{eq:M2}). But the above scaling law will hold with probability $1-\nu^{-1}$ that corresponds to the infinite regime. Let $\langle n \rangle_{pl}$ be the average outbreak size in the power-law regime. In a finite-sized system, there will be another scale $L$ up to which the power-law regime holds, and any outbreak exceeding that enters the infinite regime. Using eq.\ \ref{eq:Pn_nu_gt_1}, we can estimate the scaling in the power-law regime.
\begin{equation} \label{eq:n_avg_pl}
	\langle n \rangle_{pl} \sim \sum_{n \le L} n^{(1-\nu)/2} \sim
	\begin{cases}
	 	L^{(3-\nu)/2} & \text{$\nu \in (1,\infty) \backslash \{3\}$},\\[3pt]
	 	\log L & \text{$\nu = 3$.}
	\end{cases}
\end{equation}
$L$ can be deduced by noting that in the limit of $\nu \to 1$, the scaling law (eq.\ \ref{eq:n_avg_pl}) has to approach $N^{2/3}$ in order for the exponent to be consistent with the scaling laws for $\nu \le 1$ (eq.\ \ref{eq:M1} and \ref{eq:M2}). This is true only when $L$ scales as the following
\begin{equation} \label{eq:L_scl}
	L \sim N^{2/3}
\end{equation}
and thus we arrive at the following scaling laws,
\begin{equation} \label{eq:L}
	\langle n \rangle_{pl} \sim
	\begin{cases}
	 	N^{1-\nu/3} & \text{$\nu \in (1,\infty) \backslash \{3\}$},\\[3pt]
	 	\log N & \text{$\nu = 3$}
	\end{cases}
\end{equation} 
The scale of $N^{2/3}$ as being the upper bound of all power-laws in $P(n)$ is confirmed in simulations (see inset in figures \ref{fig:outb_dist_scaling_nu_lt_1}, \ref{fig:outb_dist_scaling_nu_eq_1} and \ref{fig:outb_dist_scaling_nu_gt_1}). The derivation in eq.\ \ref{eq:L} assumes that $L$ only depends on $N$ and not $\nu$. Intuitively, $L$ is the scale at which the power-law regime is impacted by the finiteness of the system and thus should only depend on $N$. Other model parameters only determine how fast or slow the process approaches that scale. We now summarize the finite-size scaling laws for the average outbreak size:
\begin{equation} \label{eq:avg_outb_summ}
	\langle n \rangle \sim \begin{cases}
						N^{(1+\nu)/3} & \text{$\!\!\!\!\nu < 1$,} \\[5pt]
					   	\left(\dfrac{N}{\log N}\right)^{2/3}  & \text{$\!\!\!\! \nu = 1$,} \\[15pt]
					   	\begin{Bmatrix*}[l]
						  N^{1-\nu/3} & \text{w.p.} & \nu^{-1} \\[5pt]
  						  N^{(\nu+1)/(\nu+2)} & \text{w.p.} & 1-\nu^{-1} \\
  						\end{Bmatrix*} & \text{$\!\!\!\!\nu \in (1,\infty) \backslash \{3\}$,} \\[18pt]
					   	\begin{Bmatrix*}[l]
						  \log N & \text{w.p.} & 1/3 \\[5pt]
  						  N^{4/5} & \text{w.p.} & 2/3 \\
  						\end{Bmatrix*} & \text{$\!\!\!\!\nu = 3$.}
						\end{cases}
\end{equation}
where w.p. is an abbreviation for `with probability'. The agreement of these results with stochastic simulations is shown in figure~\ref{fig:avg_outb_size}. Similarly, the summary table for the maximal outbreak size is shown below.
\begin{equation} \label{eq:max_outb_summ}
	M \sim \begin{cases}
			N^{2/3} & \text{$\nu < 1$,} \\[5pt]
			(N^2 \log N)^{1/3}  & \text{$\nu = 1$,} \\[7pt]
			N^{(\nu+1)/(\nu+2)} & \text{$\nu > 1$.}
			\end{cases}
\end{equation}

The critical point of $\nu=1$ separates the scaling behavior of $M$ into one as being a power law with fixed exponent of $2/3$ and the other as a power law with continuously varying exponent. 

For $\nu > 1$, the scaling exponent of the average outbreak size bifurcates at the value of $2/3$; the two different exponents move in opposite directions with increasing $\nu$ (compare $1-\nu/3$ with $(\nu+1)/(\nu+2)$ both of which start off from the value 2/3 as $\nu \to 1\small{+}$). A crucial insight from these results is that the average outbreak size scales as $N^\xi$ where $\xi \in (0,1) \backslash \{2/3\}$ (at $\xi=2/3$, logarithmic corrections are present). The scaling law is always sublinear as long as $\alpha =1$, i.e., there cannot occur an outbreak that scales as $\mathcal{O}(N)$ no matter how strongly the system is driven. Only for $\alpha > 1$, would there be an $\mathcal{O}(N)$ outbreak with $\nu$ having no qualitative bearing on the statistics. This is because the multiplicative nature of the supercritical BD process always dominates the constant rate of growth from external driving.

Using the above results, we can calculate the scaling window for the scaling laws, i.e., the distance from the threshold boundary within which the scaling laws are applicable~\cite{ben2012scaling}. The scaling window is a characteristic of the finite system size and shrinks to 0 in the limit of $N \to \infty$. For finite $N$, the system need not be right at the critical threshold $\alpha = 1$ for the scaling laws to be valid. Using eq.\ \ref{eq:avg_with_M} and eq.\ \ref{eq:avg_outb_summ} we obtain,
\begin{equation} \label{eq:alpha_scl}
	\lvert \alpha - 1 \rvert \sim \begin{cases}
			\quad N^{-1/3} & \text{$\nu < 1$,} \\[5pt]
			\left(\dfrac{N}{\log N} \right)^{-1/3}  & \text{$\nu = 1$,} \\[15pt]
			\begin{Bmatrix*}[l]
			  N^{-1/3} & \text{power law regime} \\[5pt]
  			  N^{-1/(\nu+2)} & \text{infinite regime} \\
  			\end{Bmatrix*} & \text{$\nu > 1$.}
			\end{cases}
\end{equation}
For a fixed $N$, the infinite regime has the largest window that grows with $\nu$.

\subsection{Outbreak duration}
\begin{figure}[tbp]
\centering
\includegraphics[width=\columnwidth]{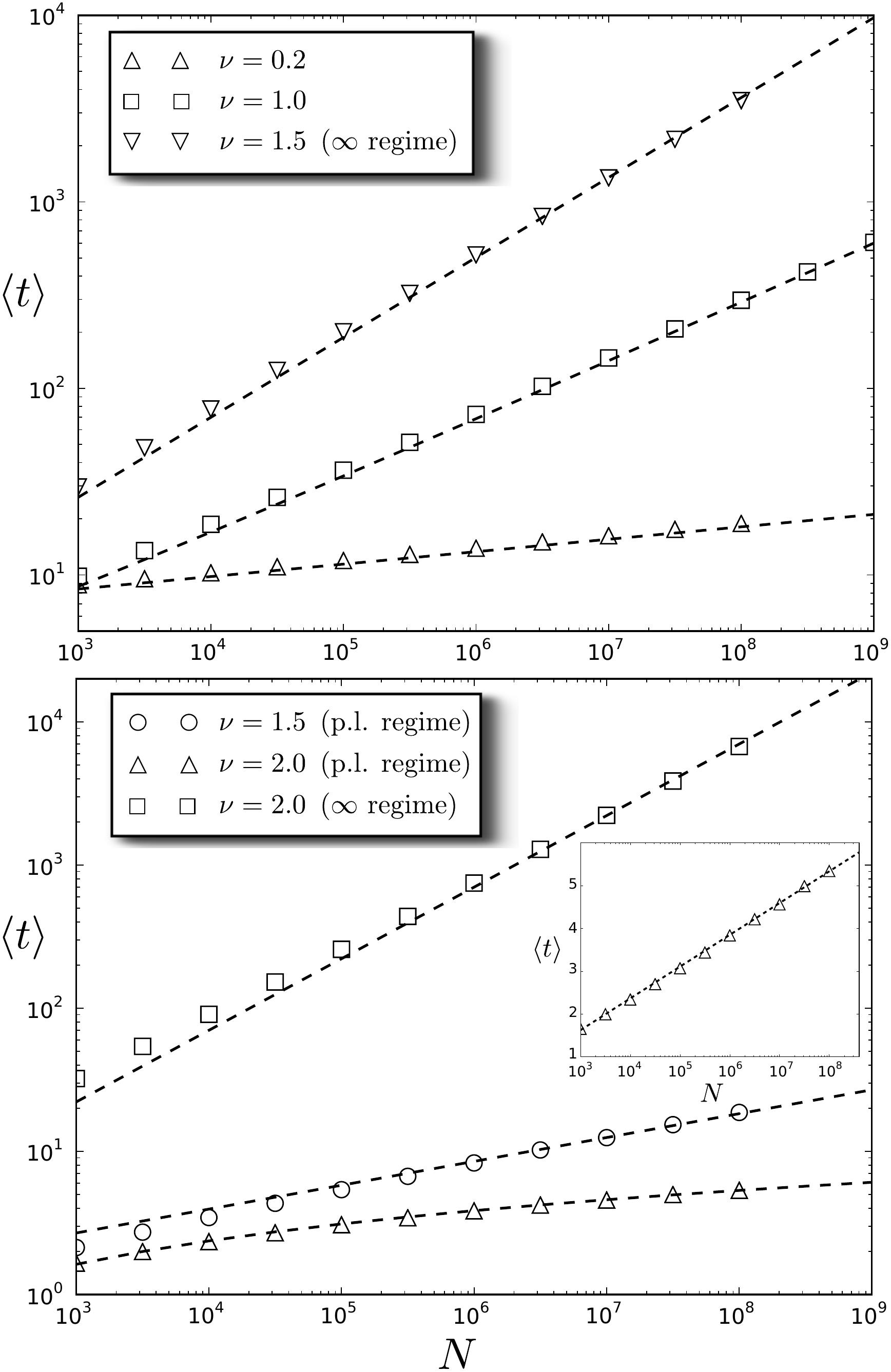}
\vspace{-.15in}
\caption{\label{fig:avg_outb_durr} Finite size scaling for average outbreak duration ($\nu=0.2,1,1.5,2$). The plots are split into two figures for clarity. The statistics of the power law regime and the infinite regime were calculated separately using $N^{2/3}$ as the separation boundary. Inset shows the scaling behavior for $\nu=2$ on a log-linear plot. Dashed lines represent scaling laws predicted from theory (eq.\ \ref{eq:avg_durr_summ}).}
\end{figure}

With the effective transmission rate $\alpha_{\star} = 1 - M/N$ below 1, the scaling behavior for outbreak durations can be obtained by using eq.\ \ref{eq:avg_dur}:
\begin{align}
	\langle t \rangle &= \dfrac{(1 - \alpha_{\star})^{-\nu} - 1}{\nu \alpha_{\star}} \notag\\
		&\sim \begin{cases}
					   	\log (N/M)  & \text{$\nu = 0$,} \\[5pt]
						\left(N/M\right)^{\nu} & \text{$\nu > 0$.}
						\end{cases}
\end{align}
For $\nu \le 1$, we arrive at the following using eq.\ \ref{eq:max_outb_summ}
\begin{equation}
	\langle t \rangle \sim \begin{cases}
						\quad \log N & \text{$\nu = 0$,} \\[5pt]
					   	\quad N^{\nu/3}  & \text{$0 < \nu < 1$,} \\
					   	\left(\dfrac{N}{\log N}\right)^{1/3} & \text{$\nu = 1$.} \\
						\end{cases}
\end{equation}
For $\nu > 1$, we have the bifurcation of behavior into the power law regime and the infinite regime. Since we already know the scale of $M$ in the infinite regime (from eq.\ \ref{eq:max_outb_summ}), we obtain
\begin{equation}
	\langle t \rangle_{\infty} \sim N^{\nu/(\nu+2)}, \quad \quad \text{$\nu > 1$.}
\end{equation}
In the power law regime, we resort to the survival function for calculating the scaling for the average outbreak duration (see Appendix \ref{sec:app_surv}) and obtain the following
\begin{equation} \label{eq:avg_vs_T}
	\langle t \rangle_{pl} \sim 
							\begin{cases}
								T_{c}^{2-\nu} & \nu \in (1,\infty) \backslash \{2\} \\[5pt]
								\log T_{c} & \nu = 2
							\end{cases}
\end{equation}
where $T_{c}$ is the cutoff timescale for the power law regime. For $\nu > 1$, we know that the cutoff length scale for the power law is $L \sim N^{2/3}$. We now estimate the relationship between $L$ and $T_{c}$.
From eq.\ \ref{eq:mean_I}, the mean number of infectious hosts increases linearly with time. Thus, the outbreak size grows quadratically with time,  
\begin{equation} \label{eq:R_vs_t}
	\dfrac{dR}{dt} \sim t, \quad R \sim t^2
\end{equation}
and this gives the relationship between $T_{c}$ and $L$ as
\begin{equation} \label{eq:L_vs_Tpl}
	T_{c} \sim \sqrt{L}
\end{equation}
The same scaling relationship was noted in \cite{ben2012scaling} for the simple SIR. Using eq.\ \ref{eq:avg_vs_T}, \ref{eq:L_vs_Tpl} and \ref{eq:L_scl}, we obtain
\begin{equation}
	T_{c} \sim N^{1/3}, \quad \langle t \rangle_{pl} \sim \begin{cases}
							N^{(2-\nu)/3} & \nu \in (1,\infty) \backslash \{2\}, \\[5pt]
							\log N & \nu = 2.
							\end{cases}
\end{equation}
The summary of the scaling laws for the average outbreak duration is given below and the agreement with stochastic simulations is shown in figure~\ref{fig:avg_outb_durr}.
\begin{equation} \label{eq:avg_durr_summ}
	\langle t \rangle \sim \begin{cases}
						\quad \log N & \text{$\!\!\!\!\nu = 0$,} \\[5pt]
						\quad N^{\nu/3} & \text{$\!\!\!\!\nu < 1$,} \\[5pt]
					   	\left(\dfrac{N}{\log N}\right)^{1/3}  & \text{$\!\!\!\! \nu = 1$,} \\[15pt]
					   	\begin{Bmatrix*}[l]
						  N^{(2-\nu)/3} & \text{w.p.} & \nu^{-1} \\[5pt]
  						  N^{\nu/(\nu+2)} & \text{w.p.} & 1-\nu^{-1} \\
  						\end{Bmatrix*} & \text{$\!\!\!\!\nu \in (1,\infty) \backslash \{2\}$,} \\[18pt]
					   	\begin{Bmatrix*}[l]
						  \log N & \text{w.p.} & 1/2 \\[5pt]
  						  N^{1/2} & \text{w.p.} & 1/2 \\
  						\end{Bmatrix*} & \text{$\!\!\!\!\nu = 2$.}
						\end{cases}
\end{equation}

On comparing the scaling laws for the average outbreak size (eq.\ \ref{eq:avg_outb_summ}) and duration (eq.\ \ref{eq:avg_durr_summ}), we note that in all power law regimes ($\nu \notin \{0,1,2,3\}$), the following relationship holds
\begin{equation}
	\dfrac{\langle n \rangle}{\langle t \rangle} \sim N^{1/3}
\end{equation}
For $\nu \in \{0,1,2,3\}$, the relationship is not too far off either with the presence of logarithmic factors,
\begin{equation} \label{eq:avg_size_by_durr}
	\dfrac{\langle n \rangle}{\langle t \rangle} \sim \begin{cases}
						\quad \dfrac{N^{1/3}}{\log N} & \text{$\quad \nu = 0$,} \\[15pt]
						\left( \dfrac{N}{\log N} \right)^{1/3} & \text{$\quad \nu = 1$,} \\[15pt]
						\quad \dfrac{N^{1/3}}{\log N} & \text{$\quad \nu = 2$,} \\[15pt]
						\quad N^{1/3} \log N & \text{$\quad \nu = 3$.}
					\end{cases}
\end{equation}
In the infinite regime however (only for $\nu > 1$),
\begin{equation}
	\dfrac{\langle n \rangle_{\infty}}{\langle t \rangle_{\infty}} \sim N^{1/(\nu+2)}
\end{equation}
the two scales converge for fixed $N$ and increasing $\nu$. This result shows that there is a universality in the power-law characteristics regardless of whether $\nu$ is below or above the critical value of 1, and it ties with the universality of $N^{2/3}$ as the characteristic scale to which all power laws extend. The presence of an `infinite regime' does not preclude this universality.

\subsection{Convergence near critical points}
The finite-size scaling laws come with a caveat: that the system size should be large enough or the parameter $\nu$ should be far away from critical points to avoid any interference from the logarithmic factors (see eq.\ \ref{eq:avg_outb_summ} and \ref{eq:avg_durr_summ}). For instance, if $\nu = 1 \pm \epsilon$, the logarithmic factor  present in the scaling laws for $\nu=1$ interferes with the scaling laws for $\nu < 1$ and $\nu > 1$ if $\epsilon \ll 1$. Since all scaling laws in the infinite regime have a monotonically increasing exponent, they would not be subjected to any interference near the critical points. Similar to eq.\ \ref{eq:alpha_scl}, we can calculate heuristically, the `interference window' for $\nu$ within which scaling power laws will be muddied via interference from logarithmic factors. Near $\nu = 1$, interference would occur if the two estimates of $\langle n \rangle$ at and above $\nu = 1$ are similar in scale, i.e.,
\begin{equation}
	\left( \dfrac{N}{\log N} \right)^{2/3} \sim N^{1 - \nu/3}
\end{equation}
which gives the interference window as
\begin{equation} \label{eq:inter_window}
	\lvert \nu - 1 \rvert \sim  \dfrac{\log (\log N)}{\log N}
\end{equation}
The window is a slowly decreasing function of $N$. The same functional form is obtained if we compare the scale of $\langle n \rangle$ at and below $\nu = 1$, as well as near all other critical points for both $\langle n \rangle$ and $\langle t \rangle$, i.e.,
\begin{equation} \label{eq:inter_window_gen}
	\lvert \nu - \nu_c \rvert \sim \dfrac{\log (\log N)}{\log N}
\end{equation}
Thus, for moderate values of $N$, the scaling laws for the power-law regime are likely to suffer from interference from logarithmic factors unless $\nu$ is far away from its critical values.

\section{Discussion}
In this work, we have solved for the statistical properties of the externally forced SIR model through rigorous analysis of the relevant stochastic process. By invoking the analogy between the BDI process and the $M/G/\infty$ queue, we were able to leverage existing results in building the theory for the process. The external driving acts as a binding agent for micro-outbreaks and is especially significant when $\alpha = 1$. In this case $\nu=1$ emerges as a second critical point in the process separating a state of recurring outbreaks from one with a single perpetual outbreak. Although power-law characteristics at the critical point were expected, the tunability of the power law by the external forcing with a precise functional form is a non-trivial result that was revealed through calculations. The finite-size scaling laws exhibit a continuum of scaling exponents governed by the driving rate that has some important implications for understanding reservoir-driven epidemics. This work also elucidates the universality of the scale of the maximal outbreak size and the ratio of average outbreak size and duration when the distribution of sizes follows a power law. 

Our results provide a framework for interpreting time series data from reservoir-driven outbreaks where the timescale of primary infections (direct reservoir transmission) and secondary infections (transmission among hosts) are comparable, and where it is not feasible to conduct field studies necessary to distinguish among them.  In cases of sufficiently weak reservoir forcing, individual chains of secondary transmission can be explained by simple SIR statistics. But we demonstrate here that if the system is near the critical threshold which is typical of emerging infectious diseases, the statistics of the process depend strongly on the reservoir forcing.  Similarly, if fine scale data were available that allowed one to resolve each micro-outbreak separately, then the simple SIR process is sufficient to describe the data.  But typically, such fine scale data are difficult to collect on the timescales of outbreaks, and practitioners often have to contend with coarse scale data on composite outbreaks, which is precisely where our theory and results serve a strong purpose.  

\begin{acknowledgments}
The authors would like to thank David Schneider, Jason Hindes and Oleg Kogan for helpful discussions.  This work was supported by the Science \& Technology Directorate, Department of Homeland Security via interagency agreement no.\ HSHQDC-10-X-00138.
\end{acknowledgments}

\appendix

\section{Derivation of generating function} \label{sec:app_pgf}
From \cite{shanbhag1966infinite}, the number of customers served in the busy period of an $M/G/\infty$ queue with arrival rate $\lambda$ and service time distribution $U(s)$ is generated by the following PGF
\begin{align} \label{eq:busy_pd}
	G(x) &= 1 - \dfrac{1}{\lambda Q(x)} \notag\\
	Q(x) &= \int_0^\infty \exp \left[ -\lambda t + \lambda x \int_0^t U(s) ds \right] dt
\end{align}
The number of customers served corresponds to the number of micro-outbreaks in the BDI process that occur on overlapping time-scales. The distribution $U(s)$ corresponds to the duration of a micro-outbreak, i.e., the BD process. The joint distribution of duration $T$ and size $R(T)$ of an outbreak in the BD process is generated by
\begin{align}
	F(0,y;s) &= \sum_{n \ge 1} \mathbb{P} \big [T \!\le\! s, R(T) \!=\! n \big ] \, y^n \notag\\
	&= \dfrac{\Lambda_0 \Lambda_1 \left(1 - e^{-\alpha (\Lambda_1-\Lambda_0) s} \right)} {\Lambda_1 - \Lambda_0 e^{-\alpha (\Lambda_1-\Lambda_0) s}}
\end{align}
where $\Lambda_0(y)$ and $\Lambda_1(y)$ are roots of the following quadratic equation such that $0 < \Lambda_0 < 1 < \Lambda_1$.
\begin{equation}
\alpha w^2 - \left(\alpha + 1 \right) w + y = 0
\end{equation}
Substituting $\nu \alpha$ for $\lambda$ and $F(0,y;s)$ for $U(s)$ in eq.\ \ref{eq:busy_pd} and simplifying the integral, we obtain the PGF for the joint distribution of number of micro-outbreaks and outbreak size in the BDI process.
\begin{equation} 
	G(x,y) = 1 - \dfrac{1}{\nu}\dfrac{\Lambda_1 \, z ^a \left(1 - z \right)^b}{\int \limits_z^1 r^{a-1} (1 - r)^{b-1} dr}
\end{equation}
where
\begin{subequations}
\begin{align*}
z = 1 - \dfrac{\Lambda_0}{\Lambda_1}, &\quad a = 1 - \nu x, \quad b = \nu \left( \dfrac{1 - \Lambda_0 x}{\Lambda_1 - \Lambda_0} \right) \\
\Lambda_0,\Lambda_1 &= \dfrac{(\alpha+1) \mp \sqrt{(\alpha+1)^2 - 4 \alpha y}}{2 \alpha}
\end{align*}
\end{subequations}

\section{Survival function for $\nu > 1$} \label{sec:app_surv}
To calculate the survival function for $\alpha=1, \nu > 1$, we repeat the calculation of the previous section but limit the integration in eq.\ \ref{eq:busy_pd} to a finite (but large) $t$ rather than $\infty$. In doing so, the PGF $G(x,y;t)$ reflects the distribution for those outbreaks that end before time $t$. 
\begin{equation} \label{eq:G_app}
	G(x,y;t) = 1 - \dfrac{1}{\nu}\dfrac{\Lambda_1 \, z_0 ^a \left(1 - z_0 \right)^b}{\int \limits_{z_0}^{z_t} r^{a-1} (1 - r)^{b-1} dr}
\end{equation}
where 
\begin{equation} \label{eq:z_t}
z_t = 1 - \dfrac{\Lambda_0}{\Lambda_1} e^{-(\Lambda_1-\Lambda_0) t}
\end{equation}
and $\alpha$ is set to 1 for $\Lambda_0$ and $\Lambda_1$. The distribution function for the duration $T$ is the total probability contained in the PGF, i.e., 
\begin{equation}
	\mathbb{P} \big [T < t \big ] = \lim_{(x,y) \to (1,1)} G(x,y;t)
\end{equation}
In the limit $(x,y) \to (1,1)$,
\begin{alignat*}{2}
	&a \to  1 - \nu, \quad 
	b \to \dfrac{\nu}{2}, \quad \Lambda_0, \Lambda_1 \to 1, \\
	&z_t \to 2 \, (1+t) \sqrt{1 - y}, \\
	&\int \limits_{z_0}^{z_t} r^{a-1} (1 - r)^{b-1} dr \to \dfrac{z_t^a - z_0^a}{a} 
\end{alignat*}
Taking the limit and simplifying the expression, we obtain
\begin{equation} \label{eq:Pt}
	\mathbb{P} \big [T < t \big ] \sim \dfrac{\nu^{-1} - (1+t)^{1-\nu}}{1 - (1+t)^{1-\nu}}
\end{equation}
As $t \to \infty$, the probability converges to $\nu^{-1}$ which is the probability that the outbreak has a finite size. With probability $1-\nu^{-1}$ the outbreak persists indefinitely. The survival function $P(t)$ is thus defined for finite size outbreaks,
\begin{align} \label{eq:Pt_asym}
	P(t) &= \mathbb{P} \big [\,t < T < \infty \big ] \notag\\
		 &\sim \dfrac{1-\nu^{-1}}{(1+t)^{\nu-1} - 1} \notag\\
		 &\sim \dfrac{1}{t^{\nu-1}} \quad \text{for large $t$}
\end{align}
The average duration can be calculated as
\begin{align}
	\langle t \rangle_{pl} \;\sim\;  -\! \int_0^{T_{c}} t \dfrac{dP}{dt} dt
					  \;\sim\;  
					  \begin{cases}
					  	T_{c}^{2-\nu} & \nu \in (1,\infty) \backslash \{2\}, \\[7pt]
					  	\log T_{c} & \nu = 2.
					  \end{cases}
\end{align}
where $T_{c}$ is a cutoff timescale in the BDI process with finite system size. We would like to note that the technique of using the survival function is also applicable when $\nu < 1$ and it yields the desired scaling laws (eq.\ \ref{eq:avg_durr_summ}) when applied.

\bibliographystyle{apsrev4-1}
\bibliography{./scaling.bib} 

\begin{thebibliography}{17}%
\makeatletter
\providecommand \@ifxundefined [1]{%
 \@ifx{#1\undefined}
}%
\providecommand \@ifnum [1]{%
 \ifnum #1\expandafter \@firstoftwo
 \else \expandafter \@secondoftwo
 \fi
}%
\providecommand \@ifx [1]{%
 \ifx #1\expandafter \@firstoftwo
 \else \expandafter \@secondoftwo
 \fi
}%
\providecommand \natexlab [1]{#1}%
\providecommand \enquote  [1]{``#1''}%
\providecommand \bibnamefont  [1]{#1}%
\providecommand \bibfnamefont [1]{#1}%
\providecommand \citenamefont [1]{#1}%
\providecommand \href@noop [0]{\@secondoftwo}%
\providecommand \href [0]{\begingroup \@sanitize@url \@href}%
\providecommand \@href[1]{\@@startlink{#1}\@@href}%
\providecommand \@@href[1]{\endgroup#1\@@endlink}%
\providecommand \@sanitize@url [0]{\catcode `\\12\catcode `\$12\catcode
  `\&12\catcode `\#12\catcode `\^12\catcode `\_12\catcode `\%12\relax}%
\providecommand \@@startlink[1]{}%
\providecommand \@@endlink[0]{}%
\providecommand \url  [0]{\begingroup\@sanitize@url \@url }%
\providecommand \@url [1]{\endgroup\@href {#1}{\urlprefix }}%
\providecommand \urlprefix  [0]{URL }%
\providecommand \Eprint [0]{\href }%
\providecommand \doibase [0]{http://dx.doi.org/}%
\providecommand \selectlanguage [0]{\@gobble}%
\providecommand \bibinfo  [0]{\@secondoftwo}%
\providecommand \bibfield  [0]{\@secondoftwo}%
\providecommand \translation [1]{[#1]}%
\providecommand \BibitemOpen [0]{}%
\providecommand \bibitemStop [0]{}%
\providecommand \bibitemNoStop [0]{.\EOS\space}%
\providecommand \EOS [0]{\spacefactor3000\relax}%
\providecommand \BibitemShut  [1]{\csname bibitem#1\endcsname}%
\let\auto@bib@innerbib\@empty
\bibitem [{\citenamefont {Andersson}\ and\ \citenamefont
  {Britton}(2000)}]{andersson2000stochastic}%
  \BibitemOpen
  \bibfield  {author} {\bibinfo {author} {\bibfnamefont {H.}~\bibnamefont
  {Andersson}}\ and\ \bibinfo {author} {\bibfnamefont {T.}~\bibnamefont
  {Britton}},\ }\href@noop {} {\emph {\bibinfo {title} {Stochastic epidemic
  models and their statistical analysis}}},\ Vol.~\bibinfo {volume} {4}\
  (\bibinfo  {publisher} {Springer New York},\ \bibinfo {year}
  {2000})\BibitemShut {NoStop}%
\bibitem [{\citenamefont {Ben-Naim}\ and\ \citenamefont
  {Krapivsky}(2012)}]{ben2012scaling}%
  \BibitemOpen
  \bibfield  {author} {\bibinfo {author} {\bibfnamefont {E.}~\bibnamefont
  {Ben-Naim}}\ and\ \bibinfo {author} {\bibfnamefont {P.}~\bibnamefont
  {Krapivsky}},\ }\href@noop {} {\bibfield  {journal} {\bibinfo  {journal} {The
  European Physical Journal B}\ }\textbf {\bibinfo {volume} {85}},\ \bibinfo
  {pages} {1} (\bibinfo {year} {2012})}\BibitemShut {NoStop}%
\bibitem [{\citenamefont {Sethna}(2006)}]{sethna2006statistical}%
  \BibitemOpen
  \bibfield  {author} {\bibinfo {author} {\bibfnamefont {J.~P.}\ \bibnamefont
  {Sethna}},\ }\href@noop {} {\emph {\bibinfo {title} {Statistical mechanics:
  entropy, order parameters, and complexity}}}\ (\bibinfo  {publisher} {Oxford
  University Press Oxford},\ \bibinfo {year} {2006})\BibitemShut {NoStop}%
\bibitem [{\citenamefont {Brauer}\ \emph {et~al.}(2008)\citenamefont {Brauer},
  \citenamefont {Van~den Driessche}, \citenamefont {Wu},\ and\ \citenamefont
  {Allen}}]{brauer2008mathematical}%
  \BibitemOpen
  \bibfield  {author} {\bibinfo {author} {\bibfnamefont {F.}~\bibnamefont
  {Brauer}}, \bibinfo {author} {\bibfnamefont {P.}~\bibnamefont {Van~den
  Driessche}}, \bibinfo {author} {\bibfnamefont {J.}~\bibnamefont {Wu}}, \ and\
  \bibinfo {author} {\bibfnamefont {L.}~\bibnamefont {Allen}},\ }\href@noop {}
  {\emph {\bibinfo {title} {{Mathematical epidemiology}}}}\ (\bibinfo
  {publisher} {Springer},\ \bibinfo {year} {2008})\BibitemShut {NoStop}%
\bibitem [{\citenamefont {Antal}\ and\ \citenamefont
  {Krapivsky}(2012)}]{antal2012outbreak}%
  \BibitemOpen
  \bibfield  {author} {\bibinfo {author} {\bibfnamefont {T.}~\bibnamefont
  {Antal}}\ and\ \bibinfo {author} {\bibfnamefont {P.}~\bibnamefont
  {Krapivsky}},\ }\href@noop {} {\bibfield  {journal} {\bibinfo  {journal}
  {Journal of Statistical Mechanics: Theory and Experiment}\ }\textbf {\bibinfo
  {volume} {2012}},\ \bibinfo {pages} {P07018} (\bibinfo {year}
  {2012})}\BibitemShut {NoStop}%
\bibitem [{\citenamefont {Lloyd-Smith}\ \emph {et~al.}(2009)\citenamefont
  {Lloyd-Smith}, \citenamefont {George}, \citenamefont {Pepin}, \citenamefont
  {Pitzer}, \citenamefont {Pulliam}, \citenamefont {Dobson}, \citenamefont
  {Hudson},\ and\ \citenamefont {Grenfell}}]{lloyd2009epidemic}%
  \BibitemOpen
  \bibfield  {author} {\bibinfo {author} {\bibfnamefont {J.}~\bibnamefont
  {Lloyd-Smith}}, \bibinfo {author} {\bibfnamefont {D.}~\bibnamefont {George}},
  \bibinfo {author} {\bibfnamefont {K.}~\bibnamefont {Pepin}}, \bibinfo
  {author} {\bibfnamefont {V.}~\bibnamefont {Pitzer}}, \bibinfo {author}
  {\bibfnamefont {J.}~\bibnamefont {Pulliam}}, \bibinfo {author} {\bibfnamefont
  {A.}~\bibnamefont {Dobson}}, \bibinfo {author} {\bibfnamefont
  {P.}~\bibnamefont {Hudson}}, \ and\ \bibinfo {author} {\bibfnamefont
  {B.}~\bibnamefont {Grenfell}},\ }\href@noop {} {\bibfield  {journal}
  {\bibinfo  {journal} {Science}\ }\textbf {\bibinfo {volume} {326}},\ \bibinfo
  {pages} {1362} (\bibinfo {year} {2009})}\BibitemShut {NoStop}%
\bibitem [{\citenamefont {Rohani}\ \emph {et~al.}(2009)\citenamefont {Rohani},
  \citenamefont {Breban}, \citenamefont {Stallknecht},\ and\ \citenamefont
  {Drake}}]{rohani2009environmental}%
  \BibitemOpen
  \bibfield  {author} {\bibinfo {author} {\bibfnamefont {P.}~\bibnamefont
  {Rohani}}, \bibinfo {author} {\bibfnamefont {R.}~\bibnamefont {Breban}},
  \bibinfo {author} {\bibfnamefont {D.~E.}\ \bibnamefont {Stallknecht}}, \ and\
  \bibinfo {author} {\bibfnamefont {J.~M.}\ \bibnamefont {Drake}},\ }\href@noop
  {} {\bibfield  {journal} {\bibinfo  {journal} {Proceedings of the National
  Academy of Sciences}\ }\textbf {\bibinfo {volume} {106}},\ \bibinfo {pages}
  {10365} (\bibinfo {year} {2009})}\BibitemShut {NoStop}%
\bibitem [{\citenamefont {Singh}\ \emph {et~al.}(2013)\citenamefont {Singh},
  \citenamefont {Schneider},\ and\ \citenamefont {Myers}}]{singh2013structure}%
  \BibitemOpen
  \bibfield  {author} {\bibinfo {author} {\bibfnamefont {S.}~\bibnamefont
  {Singh}}, \bibinfo {author} {\bibfnamefont {D.~J.}\ \bibnamefont
  {Schneider}}, \ and\ \bibinfo {author} {\bibfnamefont {C.~R.}\ \bibnamefont
  {Myers}},\ }\href@noop {} {\bibfield  {journal} {\bibinfo  {journal} {arXiv
  preprint arXiv:1307.4628}\ } (\bibinfo {year} {2013})}\BibitemShut {NoStop}%
\bibitem [{\citenamefont {Zubkov}(1972)}]{zubkov1972life}%
  \BibitemOpen
  \bibfield  {author} {\bibinfo {author} {\bibfnamefont {A.~M.}\ \bibnamefont
  {Zubkov}},\ }\href@noop {} {\bibfield  {journal} {\bibinfo  {journal} {Theory
  of Probability \& Its Applications}\ }\textbf {\bibinfo {volume} {17}},\
  \bibinfo {pages} {174} (\bibinfo {year} {1972})}\BibitemShut {NoStop}%
\bibitem [{\citenamefont {Bailey}(1990)}]{bailey1990elements}%
  \BibitemOpen
  \bibfield  {author} {\bibinfo {author} {\bibfnamefont {N.}~\bibnamefont
  {Bailey}},\ }\href@noop {} {\emph {\bibinfo {title} {{The elements of
  stochastic processes with applications to the natural sciences}}}}\ (\bibinfo
   {publisher} {Wiley-Interscience},\ \bibinfo {year} {1990})\BibitemShut
  {NoStop}%
\bibitem [{\citenamefont {Ong}(1996)}]{ong1996class}%
  \BibitemOpen
  \bibfield  {author} {\bibinfo {author} {\bibfnamefont {S.}~\bibnamefont
  {Ong}},\ }\href@noop {} {\bibfield  {journal} {\bibinfo  {journal} {Metrika}\
  }\textbf {\bibinfo {volume} {43}},\ \bibinfo {pages} {221} (\bibinfo {year}
  {1996})}\BibitemShut {NoStop}%
\bibitem [{\citenamefont {Kendall}(1953)}]{kendall1953stochastic}%
  \BibitemOpen
  \bibfield  {author} {\bibinfo {author} {\bibfnamefont {D.~G.}\ \bibnamefont
  {Kendall}},\ }\href@noop {} {\bibfield  {journal} {\bibinfo  {journal} {The
  Annals of Mathematical Statistics}\ ,\ \bibinfo {pages} {338}} (\bibinfo
  {year} {1953})}\BibitemShut {NoStop}%
\bibitem [{\citenamefont {Shanbhag}(1966)}]{shanbhag1966infinite}%
  \BibitemOpen
  \bibfield  {author} {\bibinfo {author} {\bibfnamefont {D.}~\bibnamefont
  {Shanbhag}},\ }\href@noop {} {\bibfield  {journal} {\bibinfo  {journal}
  {Journal of Applied Probability}\ }\textbf {\bibinfo {volume} {3}},\ \bibinfo
  {pages} {274} (\bibinfo {year} {1966})}\BibitemShut {NoStop}%
\bibitem [{\citenamefont {Virtamo}(2005)}]{virtamo2005queueing}%
  \BibitemOpen
  \bibfield  {author} {\bibinfo {author} {\bibfnamefont {J.}~\bibnamefont
  {Virtamo}},\ }\href@noop {} {\bibfield  {journal} {\bibinfo  {journal}
  {Lecture Notes, Helsinki University of Technology}\ } (\bibinfo {year}
  {2005})}\BibitemShut {NoStop}%
\bibitem [{\citenamefont {Press}\ \emph {et~al.}(1992)\citenamefont {Press},
  \citenamefont {Flannery}, \citenamefont {Teukolsky},\ and\ \citenamefont
  {Vetterling}}]{press1992numerical}%
  \BibitemOpen
  \bibfield  {author} {\bibinfo {author} {\bibfnamefont {W.~H.}\ \bibnamefont
  {Press}}, \bibinfo {author} {\bibfnamefont {B.~P.}\ \bibnamefont {Flannery}},
  \bibinfo {author} {\bibfnamefont {S.~A.}\ \bibnamefont {Teukolsky}}, \ and\
  \bibinfo {author} {\bibfnamefont {W.~T.}\ \bibnamefont {Vetterling}},\
  }\href@noop {} {\emph {\bibinfo {title} {Numerical Recipes in FORTRAN 77:
  Volume 1, Volume 1 of Fortran Numerical Recipes: The Art of Scientific
  Computing}}},\ Vol.~\bibinfo {volume} {1}\ (\bibinfo  {publisher} {Cambridge
  university press},\ \bibinfo {year} {1992})\BibitemShut {NoStop}%
\bibitem [{\citenamefont {Flajolet}\ and\ \citenamefont
  {Sedgewick}(2009)}]{flajolet2009analytic}%
  \BibitemOpen
  \bibfield  {author} {\bibinfo {author} {\bibfnamefont {P.}~\bibnamefont
  {Flajolet}}\ and\ \bibinfo {author} {\bibfnamefont {R.}~\bibnamefont
  {Sedgewick}},\ }\href@noop {} {\emph {\bibinfo {title} {Analytic
  combinatorics}}}\ (\bibinfo  {publisher} {cambridge University press},\
  \bibinfo {year} {2009})\BibitemShut {NoStop}%
\bibitem [{\citenamefont {Christensen}\ and\ \citenamefont
  {Moloney}(2005)}]{christensen2005complexity}%
  \BibitemOpen
  \bibfield  {author} {\bibinfo {author} {\bibfnamefont {K.}~\bibnamefont
  {Christensen}}\ and\ \bibinfo {author} {\bibfnamefont {N.~R.}\ \bibnamefont
  {Moloney}},\ }\href@noop {} {\emph {\bibinfo {title} {Complexity and
  criticality}}},\ Vol.~\bibinfo {volume} {1}\ (\bibinfo  {publisher} {Imperial
  College Press},\ \bibinfo {year} {2005})\BibitemShut {NoStop}%
\end{thebibliography}%
\end{document}